\documentclass[sn-mathphys,Numbered]{sn-jnl}

\usepackage{graphicx}%
\usepackage{multirow}%
\usepackage{textcomp}%
\usepackage{manyfoot}%

\usepackage{listings}%
\usepackage{ulem}
\usepackage{soul}
\usepackage{subcaption}
\usepackage{graphicx}

\usepackage{manyfoot}%

\usepackage{amsmath}
\usepackage{amssymb}
\usepackage{chemarr}
\usepackage{caption}

\usepackage{xcolor}
\newcommand\giada[1]{{\color{black}#1}}

\newcommand\ea[1]{{\color{black}#1}}

\newcommand{\chandra}{\textit{Chandra} \,}

\newcommand{\fermi}{\textit{Fermi-LAT}}

\newcommand{\hess}{\textit{H.E.S.S. }}

\newcommand{\fluxcgs}{erg~s$^{-1}$~cm$^{-2}$}

\newcommand{\ergs}{erg~s$^{-1}$}

\def\lsim{\;\raise0.3ex\hbox{$<$\kern-0.75em\raise-1.1ex\hbox{$\sim$}}\;}
\def\gsim{\;\raise0.3ex\hbox{$>$\kern-0.75em\raise-1.1ex\hbox{$\sim$}}\;}
\newcommand{\red}{\textcolor{black}}

\usepackage{lineno}

\begin{document}

\title[Cosmic rays]{Cosmic rays: constraints from future MeV detectors}


\author*[1]{\fnm{Giada} \sur{Peron}}\email{giada.peron@inaf.it}

\author[1,7]{\fnm{Sarah} \sur{Recchia}}\email{sarah.recchia@inaf.it}

\author[2,1]{\fnm{Stefano} \sur{Menchiari}}\email{smenchiari@iaa.es}

\author[3]{\fnm{Alexandre} \sur{Marcowith}}\email{Alexandre.Marcowith@umontpellier.fr}

\author[4]{\fnm{Andrei} \sur{Bykov}}\email{bykovandrei123@gmail.com}
\author[5,6]{\fnm{Martin} \sur{Pohl}}\email{marpohl@uni-potsdam.de}

\author[1]{\fnm{Elena} \sur{Amato}}\email{elena.amato@inaf.it}

\affil[1]{\orgdiv{Osservatorio Astrofisico di Arcetri}, \orgname{INAF}, \orgaddress{\street{Largo Enrico Fermi, 5}, \city{Florence}, \postcode{50125}, \country{Italy}}}

\affil[2]{\orgdiv{Instituto de Astrof\`{ı}sica de Andaluc\`{ı}a}, \orgname{CSIC}, \orgaddress{\street{Gta. de la Astronom\`{i}a}, \city{Genil}, \postcode{18080}, \state{Granada}, \country{Spain}}}

\affil[3]{\orgdiv{Laboratoire Univers et Particules de Montpellier}, \orgname{CNRS/Universit\'e de Montpellier}, \orgaddress{\street{Place E.Bataillon, cc072}, \city{Montpellier}, \postcode{34095},  \country{France}}}

\affil[4]{\orgdiv{https://orcid.org/0000-0003-0037-2288}}

\affil[5]{\orgdiv{Institute for Physics and Astronomy}, \orgname{University of Potsdam}, \orgaddress{\street{Karl-Liebknecht-Straße 24/25}, \city{Potsdam}, \postcode{14476}, \country{Germany}}}

\affil[6]{\orgdiv{Deutsches Elektronen-Synchrotron}, \orgname{DESY}, \orgaddress{\street{Platanenallee 6}, \city{Zeuthen}, \postcode{15738}, \country{Germany}}}

\affil[7]{\orgdiv{Institute of Nuclear Physics Polish Academy of Sciences}, \orgaddress{\street{PL-31342 Krakow}, \country{Poland}}}


\abstract{Cosmic rays are {charged energetic particles that permeate the interstellar medium. Their sizeable energy share and penetration power makes them essential players in the dynamical and chemical processes that rule Galactic evolution, such as the launching of outflows and the formation of star and planets.}
For these processes low-energy (MeV-GeV) CRs are particularly important, both because they are the most abundant and because they have the largest cross-section for ionization.  
The study of cosmic rays naturally connects with gamma-ray astronomy, as high-energy photons are the principal products of their interaction with the interstellar plasma. In this {article, after reviewing our current understanding of Galactic cosmic rays as derived from direct measurements, }
we present the state of the art regarding Galactic cosmic rays covering their direct observables, their acceleration processes and models for their propagation in the Galactic Disk. We present then an excursus on the current state of gamma-ray observations, and propose new prospects for investigating the physical properties of Galactic cosmic rays, exploiting the observational capability of future MeV missions.   }

\keywords{Low-energy cosmic rays, gamma-ray sources, MeV particles}

\maketitle

\section{Introduction}

Cosmic rays (CRs) are charged energetic particles that permeate the Galaxy and reach Earth \red{
in a nearly isotropic way \citep{abeysekara2018observation,hawc2017combined}}. Their spectral energy distribution and composition have nowadays been measured by \red{many} experiments, on balloons \cite[e.g][]{Seo2002,Yamamoto1994Balloon-borneSpectrometer}, on satellites \cite[e.g.][]{Adriani2011b,Chang2017TheMission}, on the international space station \cite[e.g.][]{Aguilar2021TheYears,Calet2020Rev}, and from the ground \cite[e.g][]{Aartsen2018AstrophysicalIceCube,Apel2013,Abreu2021TheObservatory, Lv2025PreciseComponents}, covering energies from a few GeV up to \red{more than} 10$^{20}$eV. 

CRs are mostly protons and He nuclei, with \red{a small} fraction of heavier nuclei and a very small fraction of electrons and anti-matter particles. The all-particle spectrum is generally approximated as a sequence of power laws, although, when looking at the spectra of individual species, a number of features appear, not all of which are understood \cite{Lipari2020,Recchia-2024-features-CR-transp}.
A steepening of the spectrum, known as the "knee", is detected around a few PeV and is believed to represent the maximum energy reached by protons within our Galaxy, while higher-energy CRs are believed to be heavier elements accelerated in Galactic sources up to an energy around $10^{18}$ eV \cite{augeregal20}, and particles of extra-Galactic origin beyond that energy.
This implies that the Milky Way must host CR accelerators able to reach an energy of at least a few PeV.
While the origin of CRs is still a subject of debate \cite{Gabici-2019-review}, Supernova Remnants (\red{SNRs}) are still considered as the main accelerators of Galactic CRs. However, a contribution from different types of sources (star clusters, binary systems, pulsar environments) is not excluded. In addition, it is to-date not clear, both from an observational and from a theoretical point of view \cite{2021JPlPh..87a8401A}, which source class acts as PeVatrons \cite{NatPeVatrons}, namely as factories of PeV particles. 
As far as CRs of energy below 1~GeV (hereafter Low Energy CR or LECRs) are concerned, the situation is even less clear, due to the lack of direct measurements from sources. The sources responsible for high-energy CRs are naturally expected to also produce MeV particles and, consequently, produce observational signatures. On the other hand, more source classes could be contributing in the MeV regime, without providing a sizeable contribution at GeV/TeV energies. Immediate examples are: young protostars \citep{2016A&A...590A...8P} and cataclysmic variables (CVs) \cite{2018PASJ...70...29K} \giada{, while other sources like  X-ray binaries (XRB) \cite{2022A&A...665A.145E}, normally detected in the keV domain, recently emerged also in the TeV sky, thanks to the observations of HAWC \citep{hawc_ss433,hawc_v4641} H.E.S.S. \citep{hess_s433,hess_v4641} and LHAASO \citep{lhaaso_mq} towards a handful of jetted black-hole binary systems}. This type of sources are reviewed in \citep{AmatoISSI} \red{and therefore are not discussed here.}

Even direct measurements of LECRs suffer from technological limitations: only recently, with the transit of Voyager outside the heliosphere, the lowest energy range of the CR spectrum, below a few tens of GeV could be efficiently explored \cite{Voyager1-2016,Voyager2-2019}. In that energy band, in fact, solar modulation\red{, namely the periodic change of CR intensity induced by their interaction with the magnetized Solar Wind,} plays a significant role and unbiased measurements of the CR spectrum in this energy band can be performed only in the outer interstellar medium (ISM). 
\red{These observations provide, for the first time, a direct view of the interstellar CR spectrum down to a kinetic energy per nucleon of a few MeV (see Fig.~\ref{fig:cr_spectrum} and Fig.~\ref{fig:B-over-C}), and of secondary-to-primary ratios such as B/C down to $\approx 100~$MeV/n (see Fig.~\ref{fig:B-over-C}). They are  discussed in detail in Sec.~\ref{sec:acc_trans}, in the context of a broader introduction on CR acceleration and transport.
}
Moreover, from a modeling point of view, a link between in-situ accelerated energetic particles and Earth-detected CRs is still missing, due to the complexity of the escape process from the sources and of the transport in the Galaxy.

Understanding LECRs is important for many purposes. First, MeV and sub-MeV protons and electrons can efficiently ionize matter even at high column densities, where most energetic photons cannot penetrate. In turn, ionization controls magnetic field and gas coupling and plays an essential role in star formation \cite{2020SSRv..216...29P}. Secondly, 1-100 MeV protons can produce gamma-ray line radiation in different ways: by direct excitation of nuclear states, through the production of excited nuclei, by producing radioactive species which decay through radiation, and through the production of neutrons and positrons \cite{1979ApJS...40..487R, 2003EAS.....7...79T}. Thirdly, multi-keV electrons produce X-rays due to direct Bremsstrahlung emission  and by radiative recombination lines \cite{2003EAS.....7...79T}. Finally, GeV protons are also important because of their direct imprints on gas dynamics in the interstellar medium, as these particles carry most of the non-thermal pressure. In particular, the CR pressure gradient is a force that can drive gas motion into winds from the Galactic disc. This process of gas removal from the disk contributes to lower the star formation rate in the disc \cite{2023A&ARv..31....4R}.\\

In this article, we aim at giving an overview on the state of the art regarding Galactic CR acceleration and transport mechanisms (Sec. \ref{sec:acc_trans}), with a specific focus on the low-energy component. More general reviews on the topic are provided by \citep{blasirev,amatorev,Gabici-2019-review,gabici2022}. Here we discuss the current status of gamma-ray observations of Galactic CR accelerators (Sec. \ref{sec:sources}), and propose new case studies appropriate for the next generation of MeV detectors both in terms of sources (Sec. \ref{sec:signatures}) and in relation to the diffuse medium (Sec. \ref{sec:ism_signatures}).  \ea{ The main purpose of this effort is to build a solid scientific case for the need of new instrumentation in the MeV regime to finally clarify a number of fundamental open questions related both to CR physics and to the impact of CRs on the interstellar medium chemistry and dynamics.}

\section{Cosmic-ray acceleration and transport in the Galaxy}\label{sec:acc_trans}
\begin{figure}
    \centering
    \includegraphics[width=0.5 \linewidth]{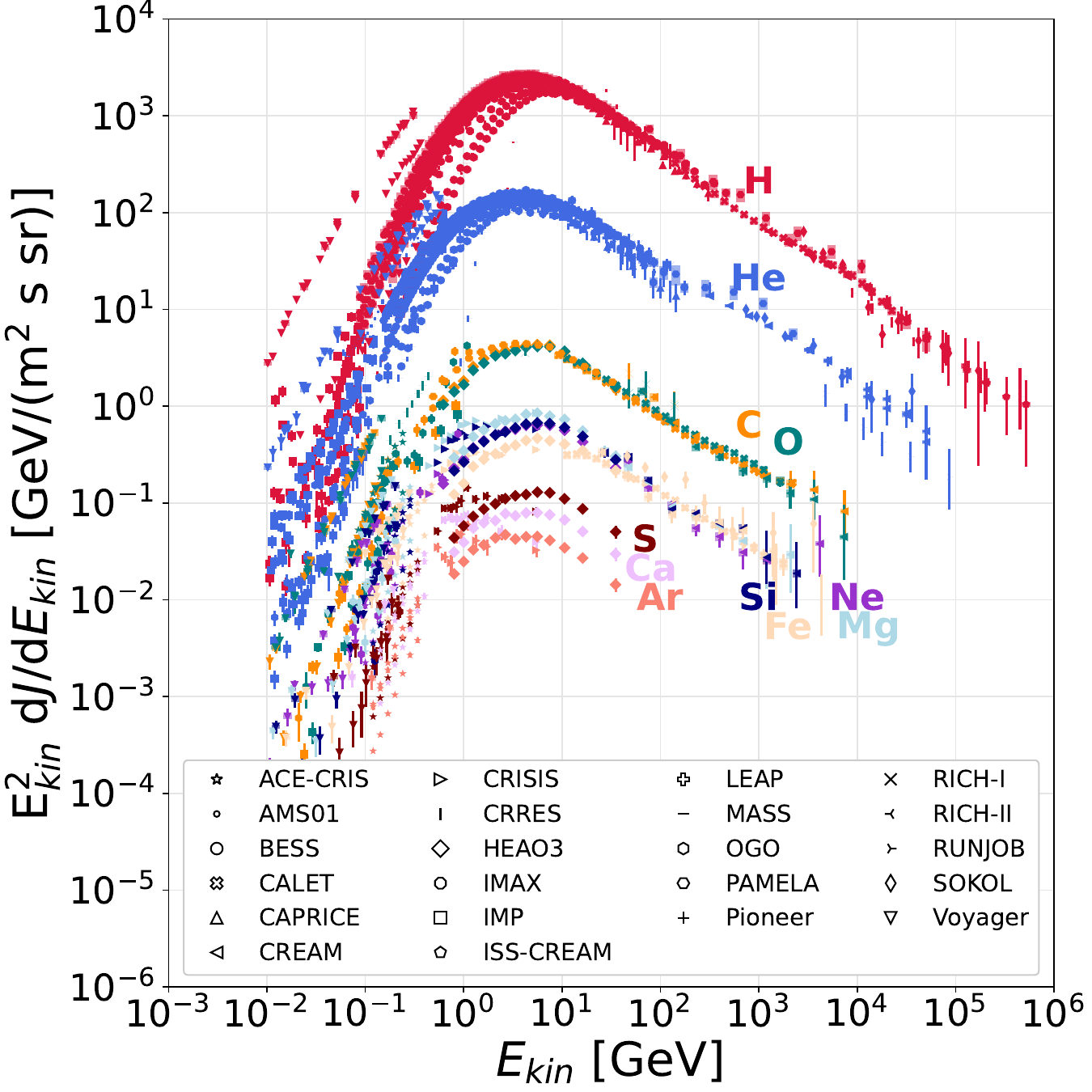}\includegraphics[width=0.5 \linewidth]{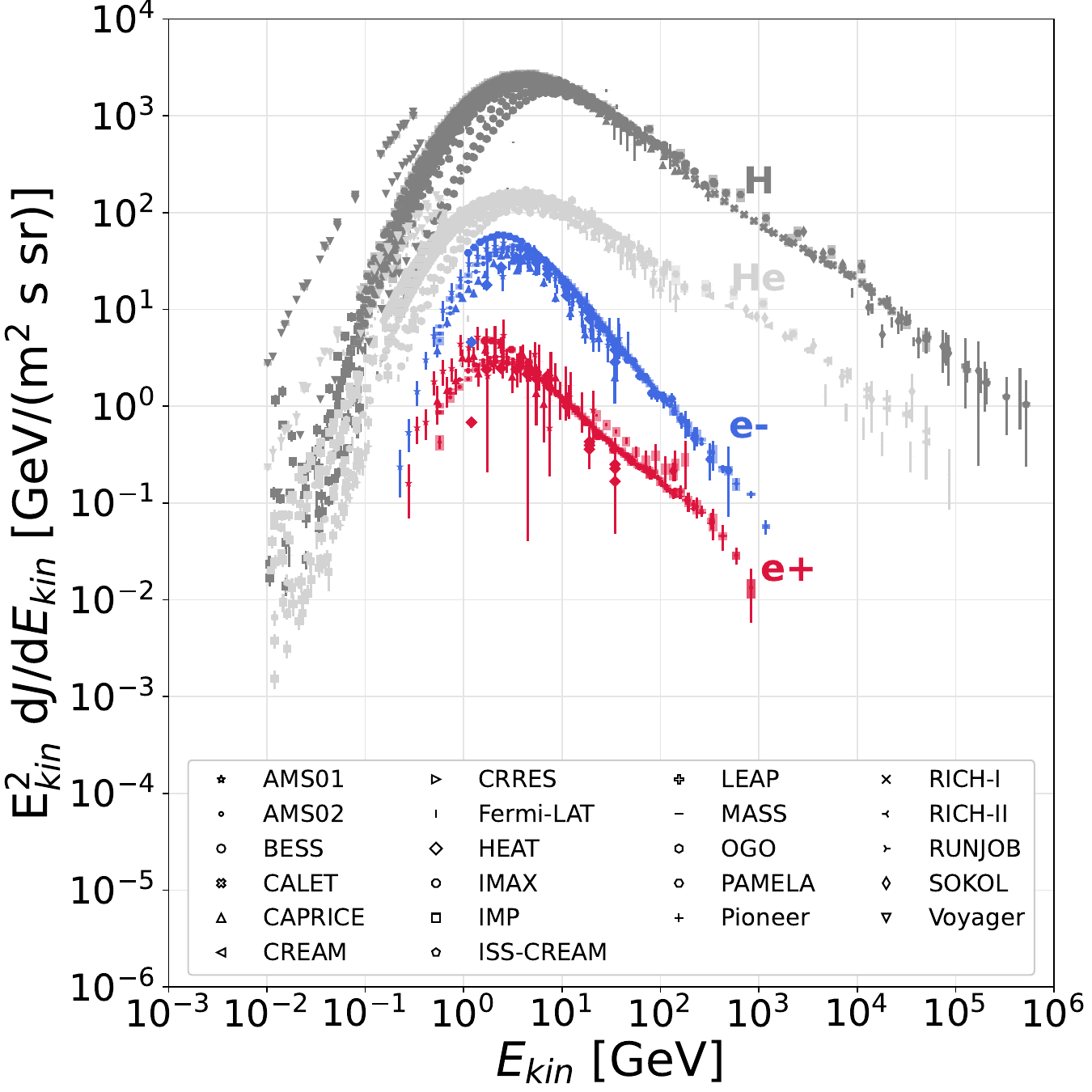}
    \caption{CR spectral energy distribution up to the knee as a function of kinetic energy for different species: on the left the spectra of nuclei are reported; on the right the spectra of leptons are plotted together with the spectrum of CR hydrogen and helium nuclei. The data were recorded by several experiments (as listed in the Figure legend) and refer to the collection of the Cosmic-Ray Database (CRDB; \citet{Maurin2014}). }
    \label{fig:cr_spectrum}
\end{figure}

\begin{figure}
    \centering
    \includegraphics[width=1\linewidth]{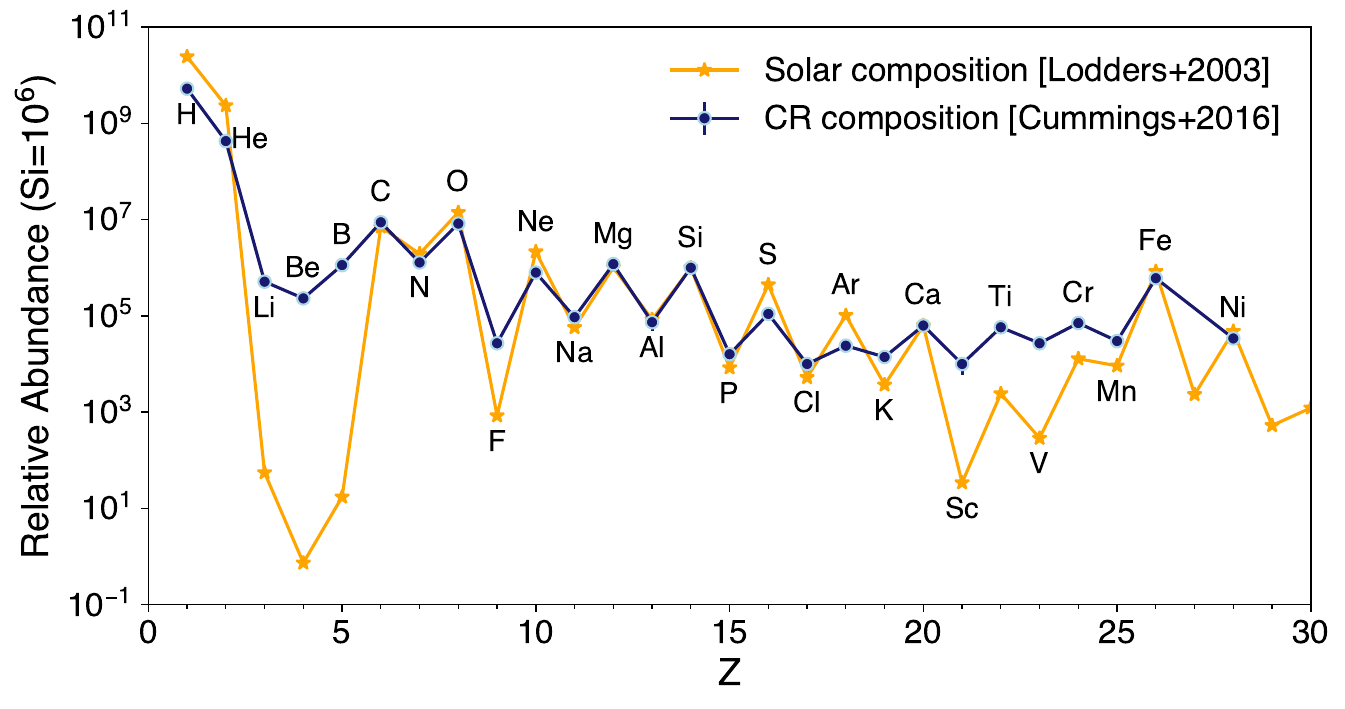}
    \caption{Abundance of elements present in the local interstellar medium (yellow line) and in the measured CRs (blue line). The \red{CR data refer to Voyager measurements at 80 MeV \citep{2016ApJ...831...18C}}.}
    \label{fig:cr_composition}
\end{figure}

The CR fluxes and flux ratios detected at Earth and in the {local Interstellar Medium (LIS)} result from a complex combination of particle acceleration in astrophysical sources and transport in the Galaxy. The collection of measurements performed by different experiments below an energy of 1 PeV is reported in Figure \ref{fig:cr_spectrum}, differentiating for the different particle species, and report to us a few important piece of information. First, we note that the most abundant element in CRs is hydrogen, but other elements of the periodic table are present in different proportions: helium constitutes 9\% of the CRs, other nuclei represent about 1\%, while leptons like electrons and positrons sum up to 2\%. The composition can be seen in Figure \ref{fig:cr_composition} and reflects the solar abundances with some noticeable exceptions in correspondence of a few elements. The most striking of these exceptions is the overabundance of lithium, beryllium, and boron and the elements below the iron peak. These are believed to originate as secondary products of spallation of interstellar CNO nuclei by CRs \citep{Reeves1970GalacticStars,Lemoine1998GalacticElements,Reeves1994OnDisplay=inlinemiZ/mimolt/momn/mnmn6/mnmn/mn/math/span}. 

Second, the CR flux measured at Earth requires a remarkably large luminosity, at the level of $\sim 10\%$ of the  \red{kinetic} power of SNe explosions in the Galaxy. Finally, the arrival direction of particles is remarkably isotropic. In addition, indirect radio, X-ray and gamma-ray observations, suggest a rather uniform distribution of CRs across the Galaxy. 

All the above-cited observations, at least to the zeroth order, are currently explained in the framework of the so-called SNR paradigm, which can be summarized  as follows \citep{DruryBook}: 
\begin{itemize}
    \item[i)] the bulk of CRs, namely nuclei of energy around $\sim 1$ GeV per nucleon, which carry most of the CR intensity, and up to larger energy (possibly up to the \textit{knee}), are accelerated at the shocks of supernova remnants, with a conversion efficiency at the level of $\sim 10\%$; 
    \item[ii)] CR particles undergo a diffusive motion in a kpc-scale magnetized halo. This motion allows CRs for multiple crossings of the Galactic disk, where rare nuclei, as Li, Be and B, can be effectively produced by the fragmentation of heavier CR nuclei. In addition, effective diffusion also explains the high level of isotropy in the particles' arrival directions. 
\end{itemize}
Despite its success in accommodating most CR data, in recent years, with the growing precision and availability of measurements and observations, this  picture has been partially questioned, as illustrated in detail in the recent review \cite{Gabici-2019-review}. Major questions are, among others, the actual capability of SNRs to accelerate particles to PeV energies \citep{Schure13,Cardillo2015OnWind,Cristofari2020,Diesing2023TheRays}, the possible contribution of other CR sources to these high energies and to the CR composition, and the detailed mechanisms of CR escape from accelerators and transport on a variety of spatial scales. 

In the following subsections, we present the key concepts related to the acceleration and propagation of Galactic CRs, with particular attention to LECRs.

\subsection{Acceleration of low-energy cosmic rays} 
The acceleration mechanism typically invoked for the production of Galactic CRs is the so-called diffusive shock acceleration (DSA), which is considered the principal process operating in SNRs and likely in other relevant astrophysical sources, such as star clusters and stellar winds, and naturally produces power-law spectra in particle rigidity, $R\equiv \frac{pc}{q}$, where $p$ is the particle momentum and $q$ its charge. 

The basic picture of DSA (see e.g. \cite{blasirev, amatorev}) can be summarized as follows. At a shock, the fluid undergoes a density and velocity jump, with the fluid speed decreasing by a factor $R_s$ between upstream and downstream. This causes the presence of an unscreened electric field at the shock transition, and every time a particle crosses the shock, it gains energy in the electric field, no matter in which direction the shock is crossed (whether from upstream to downstream or {\it viceversa}). The energy gain at each crossing is small (a fraction $(4/3)(v_s/c)(R_s-1)/R_s$ of the particle energy before crossing, with $v_s$ the shock speed) but the acceleration process relies on particle diffusion: scattering by magnetic turbulence guarantees the isotropization of the distribution function on both sides of the shock, so that particles can re-cross the shock multiple times, reaching high energies.

Due to the electromagnetic nature of the interactions involved, the particle distribution will depend only on rigidity. In addition, in the absence of a characteristic scale, namely up to energies such that the particle Larmor radius is well below the system size, the spectrum is expected to be a power-law in momentum.

 \red{
 The power-law index turns out to depend only on the shock compression ratio and reads 
 \begin{equation}\label{eq:slope-DSA-TP}
     q=3R_s/(R_s-1),
 \end{equation}
where $f(p)\propto p^{-q}$. In the case of strong shocks, $R_s=4$, this leads to a universal power-law spectrum in rigidity $f_0(p) \propto p^{-4}$, at least as long as the CRs can be considered to be test particles}.

\red{If a sizable fraction of the available energy is transferred to CRs (typically an acceleration efficiency $\gtrsim 10\%$ \cite{2010ApJ...721..886K} is required for SNRs to be the main sources of Galactic CRs), their energy and momentum flux would modify the shock \citep{Drury83}. In addition, instabilities triggered by escaping CRs, especially the Bell instability \citep{Bell04}, and the ensuing magnetic field amplification, are the favored explanation for the thin X-ray rims in young SNRs \cite{Vink2012}.
}
\red{In CR-modified shocks, a precursor is created upstream of the shock, with a continuous decrease of the plasma flow speed towards the shock and the formation of a sub-shock where the compression ratio is smaller than in the unmodified-shock case. \ea{At the same time, the energy carried away by the accelerated particles makes the shock radiative, leading to an overall compression ratio between upstream infinity and downstream that is larger than 4.} High-energy particles, being able to travel further upstream, experience a larger compression ratio than low-energy particles. In light of Eq.~\ref{eq:slope-DSA-TP}, this is expected to result in concave spectra, \ea{steep at low energies and very hard} 
at high energies \cite{2001RPPh...64..429M, blasirev}. 
} 

\red{
The hard TeV $\gamma$-ray spectra expected from efficient CR acceleration have not been observed. This may indicate a CR acceleration efficiency lower than previously thought, even though \ea{the thinness of the X-ray rims appears to highlight the associated magnetic-field amplification.} 
On the other hand, the compression ratio relevant for CRs is that of the scattering centers, namely the plasma waves that \ea{scatter particles} across the shock. When magnetic-field amplification is efficient, these waves can drift away from the shock with non-negligible speed relative to the fluid, thus reducing the effective compression ratio experienced by the particles. 
The Bell modes, that are generally thought to carry most of the magnetic-field amplification, may have a too small phase velocity \cite{Bell04}, but there is at least one simulation that does show such an effect \cite{2020ApJ...905....2C}.
}
\ea{Moving to the other extreme of the spectrum, namely the sub-GeV to GeV range, the lack of sensitive instrumentation at MeV energies prevents us from probing the particle spectral shape exactly in the region where \giada{we expect the greater divergence between a standard and a concave particle spectrum. This aspect could be tested with future MeV instruments, although the actual detection of such an effect, which predicts a spectral variation of $\Delta\alpha\lesssim 0.2$, is subject to details of particle acceleration (including the ratio between the number of protons and of electrons) and to the ambient conditions (gas, magnetic field, and radiation energy density) that determine the dominant interaction the accelerated particles are subject to. Especially the latter, are likely to change case by case. } 

}

This same spectral region is also fundamental to probe the injection of particles in the acceleration process. A major issue of any acceleration theory is the so-called injection problem, namely the problem of how thermal particles can gain enough energy to be able to start crossing the shock back and forth, instead of simply being advected away from it. 
Traditionally, a phenomenological description of injection has been adopted following the idea that particles in the tail of the Maxwellian distribution of the plasma downstream of the shock have a Larmor radius large enough to be able to leak back (thermal leakage) to the upstream region \citep{Kang-2002-injection-thermal-leakage, Blasi-2005-injection}. In this picture, the large Larmor radii of dust grains, due to their large mass-to-charge ratio, make them favoured to be injected into the acceleration process.
The subsequent destruction of the grains directly injects in the process the refractory elements, such as Mg, Si and Fe, which indeed are relatively more abundant in CRs compared to volatile elements \citep{Tatischeff-2021-composition}.
At the same time, the small mass of electrons makes them more difficult to be injected, and some level of pre-acceleration may be needed \citep{Amano-2022-injection-electrons}.

A realistic assessment of particle injection needs the study of the shock microphysics at plasma-kinetic scales, carefully \ea{accounting for the magnetic field geometry and, in particular, the field inclination relative to the shock normal}. This level of detail can only be captured through fully self-consistent kinetic plasma simulations. For instance, hybrid simulations of non-relativistic quasi-parallel shocks showed the important role played by the time-dependent shock reformation on injection, with ions impinging on the shock that may be most easily reflected when the discontinuity is steepest \citep{Caprioli-2015-injection}. Other mechanisms, such as pre-acceleration processes (particularly relevant for the injection of electrons), shock-drift and shock-surfing acceleration, and their connection to particle injection and to DSA  are also actively investigated \citep{Kumar-2021-acceleration-oblique-shocks, Amano-2022-injection-electrons, Morris-2022-pre-acceleration, Morris-2023-pre-acceleration}.\\

In addition to DSA, a mechanism that may be relevant for LECRs is the diffusive re-acceleration (second order Fermi acceleration), where CRs gain momentum when scattering on magnetic inhomogeneities in the ISM. In this process, that can be described as diffusion in momentum space, a particle with velocity $v$ can both gain (head-on collisions) or loose \red{(tail-on collision}) momentum by an amount of $\vert \delta p\vert/p \sim u_w/v$, where $u_w$ is the typical propagation velocity of the magnetic fluctuations. However, since head-on collisions are slightly more probable than tail-off collisions, particles obtain a net momentum gain on average, by $\langle \Delta p/p \rangle \sim (u_w/v)^2$. Since typical fluctuations in the ISM are largely non-relativistic, this process is quite slow, being dependent on the small quantity $(u_w/v)^2$. Thus, it can only affect the spectrum of LECRs, although its actual relevance is debated \citep{Strong-2007-CR-propagation, Thornbury-2014-reacceleration}. \red{The situation may be different in contexts where the level of turbulence and/or the wave velocity is high, such as in Galaxy Clusters \cite{Brunegtti2014-GalaxyClusters} and AGN jets \cite{Rieger2019-ShearingFlows}. These conditions are, however, probably less common in the environment of our Galaxy.}

\red{
As a final remark, it is worth emphasizing the  following point. The DSA process produces power-law spectra  in rigidity (momentum per charge), $f(p)  \propto  p^{-q}$ that are not power laws in kinetic energy, $E_\mathrm{k}$. Indeed,  the associated distribution in kinetic energy, $N(E_k)$, is related to $f(p)$ through $4\pi p^2 f(p)\, dp = N(E_k)\,dE_k$, namely
\begin{equation}
N(E_\mathrm{k}) \propto p^2 f(p) \bigg\vert \frac{dp}{dE_\mathrm{k}}\bigg\vert
\propto (E_\mathrm{k}^2 + 2\,m_0 c^2\, E_\mathrm{k})^{-\frac{q+1}{2}}\ (E_\mathrm{k} + m_0 c^2),
\end{equation}
where $m_0$ is the rest mass of the particle in question. Given the much larger mass of ions compared to electrons, below the rest-mass scale, at about $1$~GeV/n,  CR ions would have a spectrum $\propto  E_\mathrm{k}^{-\frac{q+1}{2}}$ that is harder than that of the electrons, $E_\mathrm{k}^{-q +2}$, by $(q-3)/2$ in index. Despite being far more numerous at a few GeV, CR ions may therefore be fewer than CR electrons at a few MeV.
}

\subsection{Transport of low-energy cosmic rays}
\label{subsec:lecr_transport}
\begin{figure}
    \centering
    \includegraphics[width= 0.5\textwidth]{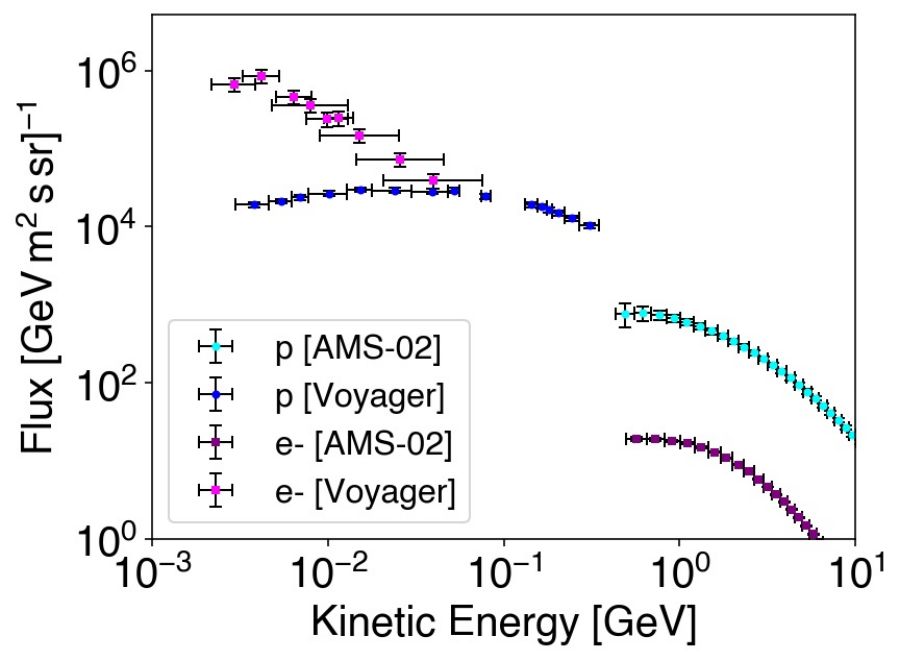}\includegraphics[width= 0.5\textwidth]{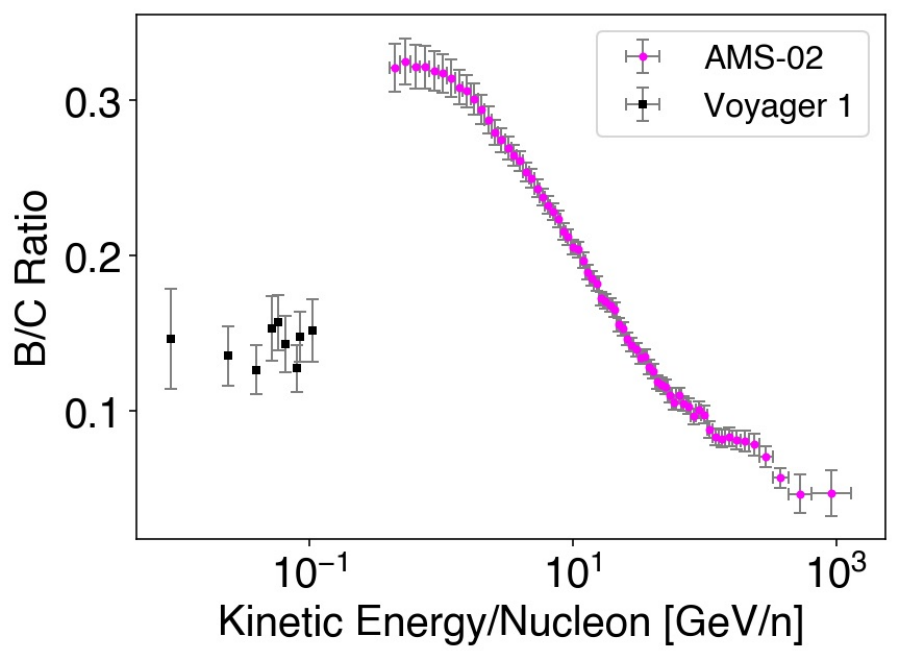}
    \caption{\red{\textit{Left panel}: Proton (blue and cyan dots) and electron (magenta and purple squares) fluxes as a function of the kinetic energy. \ea{The measurements are by AMS-02(\cite{AMS02-2016-protons-antiprotons, AMS02-2021PhR...894....1A}) and Voyager (\cite{Voyager1-2016, Voyager2-2019}), as specified in the figure inset}.}  \textit{Right panel}: Boron-over-Carbon ratio as a function of the kinetic energy per nucleon. Measurements are by Voyager (black squares \cite{Voyager1-2016}) and AMS-02 (magenta dots \cite{AMS02-2016-B-over-C}).}
    \label{fig:B-over-C}
\end{figure}

After being released from their sources, CRs undergo a tortuous journey through interstellar space, shaped by the complex interaction with the magnetized and  \red{turbulent ISM}, and eventually escape the Galaxy \citep{Amato-2018-review-transp, Gabici-2019-review}. During this time, particles occasionally interact with the interstellar gas to produce secondary CRs, most notably Li, Be and B (but also heavier nuclei, antiprotons and positrons), which are rare in the ISM but comparatively rather abundant in CRs. The secondary-to-primary ratios, such as the boron-to-carbon (B/C) ratio, provide an indication of the average amount of matter (the so-called \textit{grammage}) traversed by CRs during their propagation through the Galaxy. To a good approximation, the grammage is directly related to the residence time of CRs in the Galactic disk, namely the region where most of interstellar gas can be found. This residence time, which is measured to be at the level of $\approx$ a few Myr for GeV particles, is orders of magnitude larger than what is expected for ballistic escape of relativistic particles. In addition, the relative abundance of short-lived and stable isotopes of a given element provides an independent estimate for the total residence time of particles.  Most notably $^{10}{\rm Be}$, with a half-life of $\sim 1.4$ Myr represents a clock for the CR confinement in the Galaxy, which is found to be $\approx$ a few tens Myr, namely much larger than the residence time in the disk. To zeroth order, these measurements can be explained with CRs undergoing a diffusive motion in a kpc-scale propagation region (the Galactic halo), which extends well beyond the Galactic disk. In addition, above GeV energies, the decrease of the B/C ratio with the particle energy (see Fig.~\ref{fig:B-over-C}) can be explained with the assumption that diffusive confinement becoming less effective with increasing energy, namely with a diffusion coefficient that increases with energy as (see e.g. \cite{Gabici-2019-review})
\begin{equation}
    D(E) \approx  10^{28}-10^{29} \left(\frac{E}{\rm (a\; few)\; GeV} \right)^{0.3-0.7} \, \rm cm^2/s.
\end{equation}

Instead, the drop of the B/C ratio and its rather flat energy dependence below $\sim 1\, \rm GeV$ is currently not well understood. \\

In terms of microphysics, the CR transport is the result of the interaction of particles with the Galactic magnetic field. The latter is the superposition of a large-scale coherent structure, characterized by a complex geometry affected by the spiral arms, with components both parallel and perpendicular to the Galactic plane, and a turbulent component,
made of a variable mixture of magneto-hydrodynamic (MHD) fluctuations, both of the incompressible (Alfv\'enic) and compressible (magnetosonic) type (see \cite{Beresnyak-2019-turb-review, Schekochihin-2022-turb-review} for a review). In addition to these modes, simulations also highlighted the emergence of coherent structures and intermittency in turbulent plasmas (see e.g. \cite{Lemoine-2023-intermittency, Kempski-2023-intermittency}).
Turbulence typically originates on relatively large scales ($L_{\rm inj}\approx 1-100$ pc), as a result of random plasma motions or injection by astrophysical sources, and then undergoes a cascade process where energy is transferred to smaller scales and a turbulent spectrum is developed as a function of the wavenumber $\vec{k}$. The turbulent magnetic field is characterized by a coherence length, $L_{\rm coh}$ (typically a fraction of $L_{inj}$), the root mean square field $B_{\rm rms}$, and its relative importance compared to the mean field $B_0$, $\eta_B \equiv B_{\rm rms}/B_0$. \red{The turbulent cascade may be isotropic, as in the case of the Kolmogorov and Kraichnan-type turbulence}, or anisotropic with respect to the mean field $\vec B_0$, as in the case  of Goldreich-Sridhar turbulence \citep{Casse-Lemoine-Pelletier-2001, Mertsch-2020-test-particle-transp-review, Shalchi-2020-perp-transp-review, Lazarian-2023-review-MHD-transport}.  

Aside from large scale hydro or MHD turbulence, CRs can also trigger turbulence through a variety of particle-driven instabilities \citep{Amato-2021-stream-instability, Bykov-2013-stream-instability, Marcowith-2021-stream-instability}.

\ea{It is important to note that in the above discussion the terms \textit{Kolmogorov}, and \textit{Kraichnan} refer to the phenomenology of the turbulent cascade rather than to the ensuing properties of the turbulence spectrum or of the energy dependence of the particle scattering. 
On the other hand, the same terms, together with \textit{"Bohm"}, often indicate in CR literature a particular turbulence spectrum and a particular energy dependence of the particle diffusion coefficient, related to the former through quasi-linear theory, in spite of the fact that different cascade properties may lead to similar turbulent spectra.
In what follows, we adopt the latter convention and use the terms \textit{Kolmogorov}, \textit{Kraichnan} and \textit{Bohm} to indicate the energy dependence of the particle diffusion coefficient, $E^{1/3}$, $E^{1/2}$ and $E$, in order.}
\\

In general, the interplay of CRs with turbulent magnetic fields results in a very complex particle motion, that, on relatively small spatial and temporal scales, can be largely anisotropic with respect to the direction of the mean background field and may exhibit a non-diffusive behavior, namely a mean squared displacement in space
\begin{equation}
     \langle (\Delta x)^2 \rangle \propto  t^{\alpha},
\end{equation}
with $0 \leq \alpha \leq 2$, where $\alpha =1\,(2)$ corresponds to a diffusive (ballistic) transport. 

To illustrate, at least qualitatively, this rich phenomenology, let us follow the journey of a particle population after its release from a given source.  On scales smaller than a few $L_{\rm coh}$ from the injection site, particle would initially undergo a simple gyromotion along the mean field. However,  the interaction of CRs with turbulent fluctuations can lead to effective pitch-angle scattering along field lines, which eventually translates into spatial diffusion along field lines, characterized by a \textit{parallel} diffusion coefficient $D_{\parallel}(r_L(E))$.
Typically,  this parameter depends on the particle energy through the Larmor radius, the so-called resonant scale. Indeed, isotropic turbulence at the resonant scale $ k \sim 1/r_L$, if present, is typically the most effective agent in scattering CRs and $D_{\parallel}$ will be determined by  the spectral properties of the turbulence at the resonant scale. When $\delta B/B_0 \ll 1$, the particle transport can be treated in the quasi-linear approximation (see e.g. \cite{Mertsch-2020-test-particle-transp-review} for a review) and $D_{\parallel}$ can be expressed as 
\begin{equation}\label{eq:D-QLT}
D_{\parallel}(E)=\left.\frac{4\pi c\,r_{\rm L}(E)}{3 I(k_{\rm res})}\right|_{k_{\rm res}=1/r_{\rm L}},
\end{equation}
where  $I(k_{\rm res}) = \delta B(k_{\rm res})^2/ B_0^2$ is the normalized wave energy density calculated at the resonant wavenumber. 

The effectiveness of resonant scattering depends on whether turbulence injected on scales $\gg r_L$ is able to cascade to resonant scales (typically $\ll 10^{-6}$ pc for sub-GeV particles) without 
being damped or becoming inefficient for scattering, as in the case of the anisotropic cascade of  Alfv\'enic turbulence \citep{Fornieri-2021-MHD, Lazarian-2023-review-MHD-transport}. Relevant damping mechanisms in the ISM are the ubiquitous non-linear Landau damping, turbulent damping and ion-neutral damping in partially ionized media (see e.g. \citep{Recchia-2022-grammage, Chernyshov-2023-NLLD, Cerri-2024-turbulent-damping} and references therein).

More recently, non-resonant interactions, such as trapping in magnetic mirrors produced by compressive perturbations on scales $\gg r_L$, were also found to play a role in scattering CRs \citep{Chandran-2000-cloud-mirror, Lazarian-Xu-2021-mirror-diff, Fornieri-2021-MHD, Lazarian-2023-review-MHD-transport}, together with coherent structures and intermittency \citep{Lemoine-2023-intermittency, Kempski-2023-intermittency}.\\

Taking \red{into} account this variety of effects, some authors (see e.g. \citep{Schlickeiser-2010-mean-free-path}) suggested  a scaling for the diffusion coefficient
\begin{equation}\label{eq:D-parallel-beta-gamma}
    D_{\parallel} (E)\propto \beta(E)\gamma(E)^{\delta},
\end{equation}
where $\beta = v/c$  is the ratio between the particle speed and the speed of light and $\gamma$ is the particle Lorentz factor. Above multi-GeV energies one recovers the $E^{\delta}$ scaling with a diffusion coefficient at the level of that reported in Eq.~\ref{eq:DiffCoeff}, while at low energies one finds $D_{\parallel} \propto \beta $ and a typical mean free path at the level $\lambda_{\parallel} \approx 10^{12}/\beta \,\rm cm$ can be obtained \citep{Lazarian-Xu-2021-mirror-diff}.\\

Together with turbulence of extrinsic origin, CRs can also excite plasma instabilities that can substantially enhance their
scattering along $B_0$. Most notably, the excitation of resonant Alfv\'en waves (the so-called resonant streaming instability) is triggered by gradients in the CR density \citep{Skilling-1971, Amato-2021-stream-instability}, typically found in the proximity of accelerators or between the Galactic disk and halo. This mechanism can dominate the particle transport in the Galaxy up to hundreds of GeV \cite{Aloisio-Balsi-2013-self} and may produce regions of suppressed diffusion around SNRs both for particles above (see e.g. \citep{Recchia-2022-grammage} and references therein) and below \citep{Jacobs-2022-selfconf-lowE} GeV energies.

While moving and being scattered along field lines, CR particles are also substantially affected by the random wandering of field lines, which results from turbulence on scales $\gg r_L$. In particular, the particle motion perpendicular to the mean field is a complex combination of cross-field diffusion and random motion of the field lines, where the latter is due to large scale turbulence, while the former, that allows CRs to jump from one field line to another (see \citep{Shalchi-2020-perp-transp-review} for a review), is due to a combination of scattering and drifts that result in a diffusion coefficient much smaller than $D_{\parallel}$.  
Due to the exponential divergence of neighboring field lines, the combination of the random motion of field lines and small-scale cross-field diffusion, eventually leads, on scales $\gtrsim L_{\rm coh}$ and on sufficiently long timescales \citep{Rechester-Rosenbluth-1978, Chandran-2000-cloud-mirror, Casse-Lemoine-Pelletier-2001}, to a CR perpendicular diffusion characterized by 
 $D_\perp \lesssim D_\parallel$, with the ratio tending to unity as the turbulence level increases.

Although the effects of anisotropic transport may be very relevant in the proximity of CRs sources,
on large spatial scales, namely $\gg L_{\rm coh}$,     particles traverse many uncorrelated magnetic field patches and their motion can be approximated as an isotropic diffusion in space with a global diffusion coefficient $D \sim D_{\parallel}$. Quantities such as the B/C ratio are sensitive to such global diffusion coefficient, averaged over the kpc-scale Galactic halo (see e.g. \cite{Chandran-2000-cloud-mirror, Nava-Gabici-2013-anisotropic, Recchia-2024-features-CR-transp} and references therein).\\

CR scattering not only makes the particle motion diffusive and their distribution function very close to isotropic, but also provides an effective coupling between the high-energy particles and the plasma. Due to this coupling, the CR transport is also affected by the advective motion induced by plasma waves themselves (the scattering centers) or by large-scale flows in the ISM. This coupling can also allow for CRs to effectively impart momentum and energy to the background plasma, with possible prominent effects on the ISM, such as the launching of Galactic winds (see e.g. \citep{Recchia-2020-review-winds} for a review). \\

The diffusive-advective propagation of CRs is encompassed in the transport equation
\begin{equation}\label{eq:CR-transport-general}
\frac{\partial f}{\partial t} - 
\vec{\nabla}\cdot\left[D\vec{\nabla}f\right] +\vec{u}\cdot\vec{\nabla}f - \vec{\nabla}
\cdot \vec{u}\frac{p}{3}\frac{\partial f}{\partial p} + \frac{1}{p^2} \frac{\partial}{\partial p} \left[ \dot{p}p^2 f\right] = Q(\vec{r}, p),
\end{equation}
where $D$ is the \textit{global} diffusion coefficient illustrated above, $\vec{u}$ is the advection velocity (typically assumed to be at the level of $\approx 10\, \rm km/s$, and including both large-scale outflows and advection with plasma waves) and $Q$ is the injection spectrum associated to CR sources. The $\vec{\nabla}\cdot \vec{u}$ term encompasses adiabatic losses (gain) due to the expansion (compression) of the advecting flow, while $\dot{p}$ is the momentum loss rate. 

\begin{figure}
    \centering
    \includegraphics[width=0.5 \linewidth]{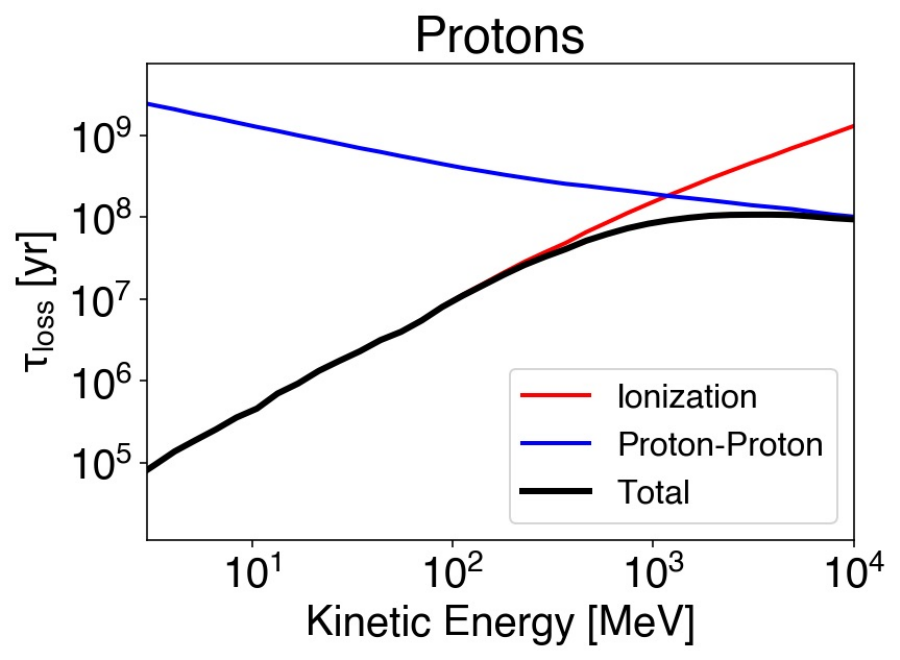}\includegraphics[width=0.5 \linewidth]{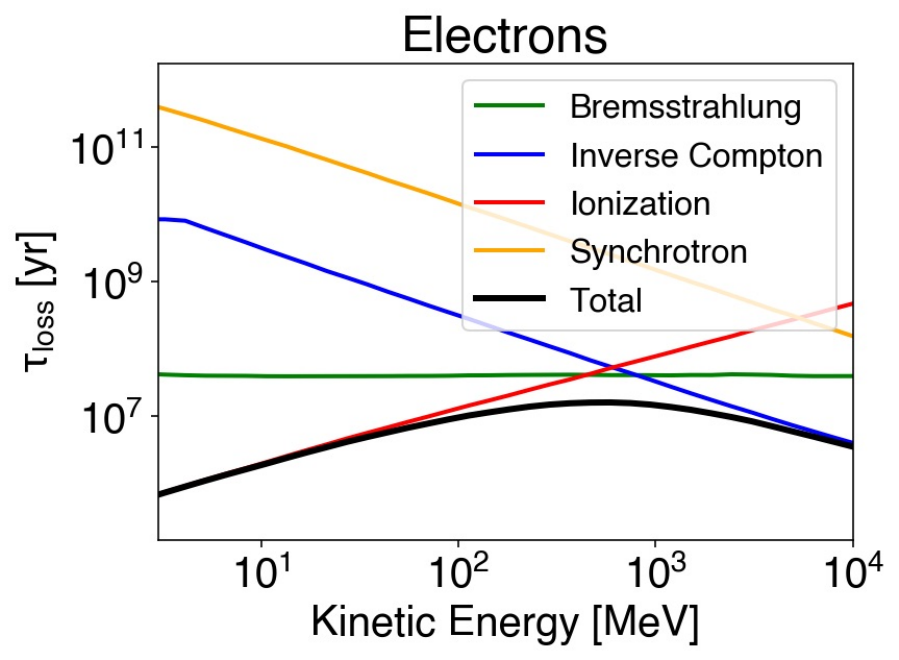}
    \caption{Typical CR loss timescales in the interstellar medium as a function of the particle kinetic energy \protect~\citep{Ravikularaman-2025-ionization-GC}.  \textbf{Protons} (left panel): ionization (red solid curve), proton-proton interactions (blue solid curve) and total (black solid curve) for a reference hydrogen density of $n_H =\, \rm  1~cm^{-3}$. \textbf{Electrons} (right panel): ionization (red solid curve) and Bremsstrahlung  (green solid curve) for a reference hydrogen density of $n_H =\, \rm  1~cm^{-3}$; inverse Compton scattering in the interstellar radiation field (blue solid curve);  synchrotron in a $B_0 = \rm 3~\mu G$ magnetic field;  total (black solid curve).}
    \label{fig:loss-time-p-e}
\end{figure}

In the case of LECRs momentum losses are due to a variety of particle interactions with the gas, with the interstellar magnetic field and photon field.
In Fig.~\ref{fig:loss-time-p-e} we report approximate values for the loss time of protons (left panel), as due to ionization and proton-proton interactions, for a reference hydrogen density of $n_H = 1\, \rm cm^{-3}$, and electrons (right panel) as due to ionization and Bremsstrahlung losses in the same reference density, inverse Compton scattering (ICS) in the interstellar radiation field and synchrotron in a $B_0 = 3\, \rm \mu G$ field (see \citep{Pohl-1993, Mannheim-Schlickeiser-1994, Padovani_MCIon_2009, Phan-2018-ionization-MC, Ravikularaman-2025-ionization-GC} and references therein). 

For energies below a few GeV, energy losses are typically dominated by mechanisms that depend solely on the gas density. This is the case for both the warm ionized phases of the ISM (typical gas density $n_H \sim 1\, \rm cm^{-3}$) and the dense and neutral environments of MCs (typical gas density $n_H \sim 100\, \rm cm^{-3}$), where 
ionization-Coulomb losses dominate below a few GeV for nuclei and below $\sim$ a few $100$ MeV 
for electrons. For the latter, Bremsstrahlung dominates from a few GeV down to $\sim$ a few $100$ MeV.\\

Given the relatively short loss timescale of sub-GeV particles as compared to the Myr scale residence time in the Galactic disk inferred from CR measurements, one may expect that the steady-state LECRs resulting from neglecting all terms in Eq.~\ref{eq:CR-transport-general}, except for injection and losses, could reproduce the spectrum measured by Voyager 1 and 2 \citep{Voyager1-2016, Voyager2-2019}. This  approach was discussed in detail in the review paper \cite{Tatischeff-Gabici-2018-review-spallogenic}, where the authors also included the effect of  CR escape from the Galactic disk in the form of particles confined in a box and having a  constant probability per unit time to escape from it (the so-called \textit{leaky-box} model). 
The resulting equation for primary nuclei, expressed in terms of the particle kinetic energy per nucleon, $E_{k/n}$, rather than in terms of momentum $p$ (given that the kinetic energy per nucleon is conserved in spallation of nuclei on the ISM), reads
\begin{equation}
    \frac{N_i(E_{k/n})}{\tau_i^{\rm tot}(E_{k/n})} + \frac{\partial}{\partial E_{k/n}}\left[\dot{E}_{k/n, i} N_i(E_{k/n}) \right] = Q_i(E_{k/n}).
\end{equation}
Here $\tau_i^{\rm tot}(E_{k/n})$ is the timescale resulting from the combination of escape and fragmentation, and the CR distribution as a function of $p$ and that as a function of $E_{k/n}$ are related to one another through
$4\pi p^2 f_i(p)\, dp = N_i(E_{k/n})\,dE_{k/n}$. This equation can be solved analytically, yielding 
\begin{equation}
    N_i(E_{k/n}) = \frac{1}{\dot{E}_{k/n, i}(E_{k/n})} \int_{E_{k/n}}^{\infty} dE\, Q_i(E) \exp\left[-\int_{E_{k/n}}^{E} \frac{dE'}{\dot{E}_{k/n, i}(E')\tau_i^{\rm tot}(E')} \right], 
\end{equation}
which, for long $\tau_i^{\rm tot}$, namely for the case in which the Galactic disk behaves as a calorimeter for LECRs, reduces to
\begin{equation}
     N_i(E_{k/n}) = \frac{1}{\dot{E}_{k/n, i}(E_{k/n})} \int_{E_{k/n}}^{\infty} dE\, Q_i(E).
\end{equation}
Using this approach, it was shown \citep{Tatischeff-Gabici-2018-review-spallogenic} that it is not generally possible to fit simultaneously the Voyager and AMS-02 spectra of nuclei and electrons by assuming an injection spectrum that is a pure power law in momentum, and that a break should be included at $\sim 200 $ MeV per nucleon and with a low-energy slope that is typically different for different nuclei. Although a low-energy break is expected in most models of particle acceleration, the possible origin of different low-energy slopes remains unexplained.\\

\begin{figure}
    \centering
    \includegraphics[width= 0.7\linewidth]{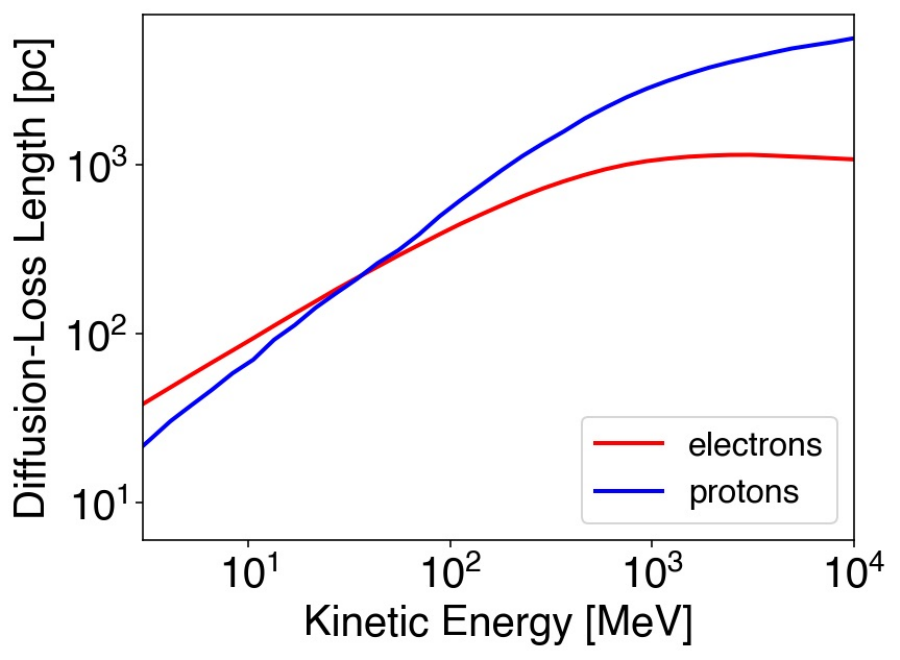}
    \caption{Typical diffusion-loss length of  CR protons (blue curve) and electrons (red curve) corresponding to the loss timescales  shown in Fig.\protect~\ref{fig:loss-time-p-e}, and with a diffusion coefficient in the form of Eq.~\ref{eq:D-parallel-beta-gamma} with $D(10\, {\rm GeV})  = 5\times 10^{28}\, \rm cm^2/s$ and $\delta = 0.63$ \citep{Phan-2023-stochasticity}.}
    \label{fig:diff-loss-length}
\end{figure}

On the other hand, the severe energy losses experienced by LECRs below a few hundred MeV significantly reduce the distance CRs can travel after being released from sources. In Fig.~\ref{fig:diff-loss-length}, we show the typical diffusion-loss length, $R_{\rm diff-loss}(E_k) = \sqrt{D(E_k)\, \tau_{\rm loss}(E_k)}$, of CR protons and electrons in a warm ionized medium, for which the corresponding loss timescales are shown in Fig.~\ref{fig:loss-time-p-e}, and with a diffusion coefficient in the form of Eq.~\ref{eq:D-parallel-beta-gamma} with $D(10\, {\rm GeV})  = 5\times 10^{28}\, \rm cm^2/s$ and $\delta = 0.63$ \citep{Phan-2023-stochasticity}. Here, for simplicity, we do not take into account that high-energy particles, which can exit from the disk and propagate in the Galactic halo, would experience a lower average gas density, which would increase their diffusion-loss length. For CRs of energy
below $\sim$ 100 MeV, the diffusion-loss length becomes comparable with the size of the disk and rapidly drops below $\sim 100$ pc for lower energies. 
In this scenario, the conventional assumption of a smooth and continuous source distribution would break down for LECRs, and the discrete nature of their sources becomes significant for interpreting  Voyager data and in modeling the ionization of molecular clouds \citep{Phan-2021-stochasticity, Phan-2023-stochasticity}. 

Indeed, when the propagation length becomes comparable to, or smaller than, the average distance between sources, the particle flux at a given location is strongly affected by the actual location and age of the sources. Since the latter are typically not well constrained, source catalogs are often partial, and there are uncertainties on the spectrum released by individual sources, the injection from a source population must be treated statistically. 
Correspondingly, the total flux, $\psi_i$, contributed at Earth by all sources of a given realization $i$ should be treated as a stochastic variable. The solution of the steady-state transport equation, Eq.~\ref{eq:CR-transport-general}, with the injection term given by the smooth source distribution (from which the individual sources are extracted), corresponds to the expectation value, $\langle \psi \rangle$, for the total flux that would be obtained by averaging over many realizations. 
Above a few hundreds GeV, the flux of each realization is typically very close to this mean flux, which, in turn, is the one that should match the data. 

However, at low energies, the statistical distribution of these fluxes is far from Gaussian. Indeed, considering a single value for the particle kinetic energy $E_k$, the possible presence of a few young and nearby sources leads to a long high-flux power-law tail in the distribution of fluxes, that may make the second moment (variance) divergent. 
As a result, the median flux - the value around which most realizations cluster — can be substantially different from the mean flux, and, as shown in detail by \citep{Phan-2021-stochasticity}, it constitutes a better representation of the typical CR flux in a given region. 

\begin{figure}
    \centering
    \includegraphics[width= 0.7\linewidth]{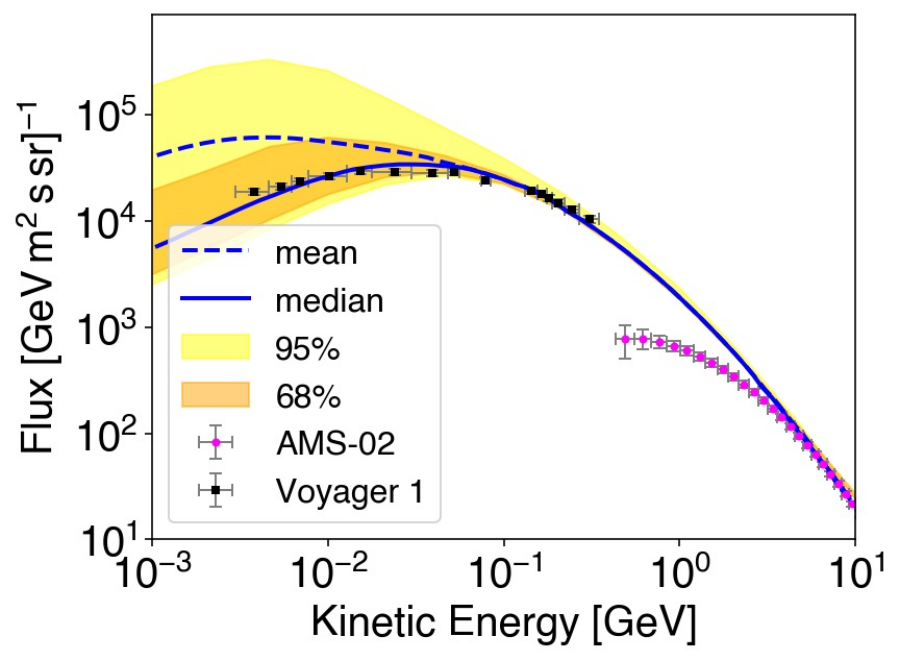}
    \caption{Stochastic fluctuations of low energy CR protons (plot adapted from \cite{Phan-2021-stochasticity}). Particles are injected with a single power-law in momentum. The stochastic median (blue solid curve)  the mean (blue dashed curve) fluxes are quite different at low energies, as a result of the non Gaussian distribution of fluxes. The orange (yellow) band represents  the $68\%$ ($95\%$) percentile bands around the median. The median flux  can roughly reproduce both the Voyager 1 \protect~\citep{Voyager1-2016} and AMS-02 \protect~\citep{AMS02-2016-protons-antiprotons} data.}
    \label{fig:stochasticity-proton}
\end{figure}

The relevance of the stochastic nature of sources on the local  LECR spectrum can be appreciated in Fig.~\ref{fig:stochasticity-proton}, where it is shown, for CR protons injected with a single power-law in momentum, the median and mean flux, and the $68\%$ and $95\%$ percentile bands around the median, for a case that can roughly reproduce both the Voyager 1 and AMS-02 data (see \citep{Phan-2021-stochasticity} for details). The sharp decrease of the diffusion-loss length at low energies (see Fig.~\ref{fig:diff-loss-length}) reflects on a sharp increase of the level of fluctuations around the median. The latter has a shape that matches well the data without the need for a break in the injection spectrum, and most realizations are distributed around it. In contrast, the shape of the mean flux would not match that of the data without including a break in the injection spectrum (in agreement with what \ea{was} found 
by \cite{Tatischeff-Gabici-2018-review-spallogenic} and illustrated above), and the mean flux lies outside of the $68\%$ uncertainty band.  This uncertainty in the flux of LECRs in different locations in the Galaxy may be a possible explanation for the level and dispersion of the ionization rate measured in MCs (see Sec.~\ref{subsec:MCIonization}) and may also consistently affect the low energy gamma-ray emission of these objects. Prospects for probing the MeV flux of clouds are discussed in \ref{subsec:MC}. These features are not unique of LECRs and may appear whenever energy losses substantially reduce the propagation distance, as it is the case for multi-TeV CR electrons \citep{Recchia-2019-local-TeVatron, Evoli-2021-stochasticity}. 
In turn, the determination of the ionization rate at different locations (for instance in active star formation regions versus inactive ones) is another way to probe the LECR content in the ISM.

\section{Sources of Galactic cosmic rays} \label{sec:sources}
Several types of source in the Galaxy meet the requirement for CR acceleration. In correspondence of such sources, gamma-ray emission is expected as a result of the interaction of freshly accelerated particles with their surrounding medium. Accelerated protons and nuclei \red{interact via} inelastic collisions with the nuclei in the ISM producing unstable particles (mainly mesons like pions, but also other particles (see \cite{ParticleDataGroup:2024cfk}) that in turn quickly decay into gamma-ray photons, which carry approximately 10\% of the energy of the parent nucleus. 

Electrons and positrons interact with the interstellar radiation, gas, and magnetic fields via inverse Compton scattering, bremsstrahlung, and synchrotron processes. In particular bremsstrahlung in the MeV band is a direct probe of MeV-GeV electrons in the Galaxy and, since the radiative efficiency is known, it is also a measure of the electron-induced heating and ionization of interstellar gas. All these radiation processes will be referred to hereafter as hadronic and leptonic emission mechanisms, respectively. In Sec.~\ref{sec:signatures}, we discuss in detail the  challenges in disentangling the hadronic from leptonic emission in CR sources, and the constraints that can be derived on the acceleration process, \ea{its efficiency and, most notably, the electron-to-proton ratio}

Gamma-ray observations are widely used to investigate the spectral properties of CRs in correspondence of their accelerators. Data collected \giada{at high (100 MeV-1 GeV), very-high (100 GeV-100 TeV), and  ultra-high (UHE; $>$ 100 TeV) energies} by the available gamma-ray instruments unveiled a variety of powerful emitters in the Milky Way: Supernova Remnants (SNRs), Pulsars (PSRs) and their nebulae (PWNe), Stellar clusters (SCs) and stellar binary systems (BIN) including micro-quasars (MQs) \citep{lhaaso_mq}. \giada{However, it is not clear if \ea{all these source classes contribute sizably to the galactic CR spectrum and in which energy band.
\red{SNR} observations from GeV to TeV energies confirmed that they are the primary contributors to the Galactic CR population, but they might not be the dominant component at energies higher than hundreds TeV. Even though UHE emission has been recorded in correspondence of a couple of SNRs \citep{lhaaso_w51,lhaaso_ic433}, the interpretation of the emission is challenged by the vicinity of other sources, like SCs and PWNe.} PSR environments, and especially PWNe, have been confirmed as strong emitters in all observed energy bands. However they are likely accelerating mostly leptons, and while they are probably the primary CR positron factories in the Galaxy \citep{evoli2021} their contribution in terms of hadrons, which are the main CR constituents, is only marginal.  }
We refer to \citep{AmatoISSI} for a discussion about these objects and their role in the MeV domain. \giada{Similarly, the role of MQs as CR accelerators is still debated: on one hand, there are strong hints in favour of a leptonic interpretation of the emission of  the microquasars SS~433 and V4641~Sgr, like the absence of dense gas regions in their surroundings and an apparent energy-dependent morphology \citep{hess_s433,hess_v4641}; on the other hand, a fraction of hadrons is expected, and is invoked to explain the highest energy part of the spectrum \citep{lhaaso_mq}. A comprehensive discussion about the latter source class however is beyond the scope of this review, and we refer to \citep{AmatoISSI}. }

\giada{In the following sub-sections, we focus on SNRs (Sec.~\ref{subsec:SNR}) and SCs (Sec.~\ref{subsec:SC}) and} we describe the \giada{currently available observations in the GeV-TeV energy range and try to forecast what are the most promising sources to observe in the MeV. The physical constraints that can be derived with such MeV observations on these sources are later discussed in Sec.~\ref{sec:signatures}. }
 
\subsection{Supernova remnants}\label{subsec:SNR}
\red{SNRs} are believed to be the main responsible for CR acceleration in the Galaxy. This idea is supported by three main arguments: i) the energetics provided by supernova explosions with the standard supernova event rate (3 explosions per century) is sufficient to supply the power of the observed population of CRs; ii) the environment inside a remnant provides the conditions for diffusive shock acceleration (see Sec. \ref{sec:acc_trans}); iii) supernova remnants are bright in gamma-rays, in the GeV-TeV range. 
Nevertheless, several questions remain unsolved about the contribution of SNRs to the entire Galactic population of CRs and could be more deeply investigated with MeV observations. First, in order to connect the gamma-ray observations of SNRs to the CRs observed at Earth, it is of fundamental importance to understand whether CR protons are responsible for the observed gamma-ray emission, rather than leptons; second, even in the cases where hadronic emission mechanisms dominate, it is fundamental to understand whether leptons are also present and in what fraction. This information is essential for a correct assessment of the energetics of the acceleration inside these sources, and especially to test if they could reach the highest energy end of the Galactic CR spectrum.

{While it has been known for decades that in order for SNRs to accelerate particles to the knee efficient magnetic field amplification (MFA; \giada{ see Section \ref{sec:acc_trans}}) is required \citep{schure2012,amatorev,blasirev}, the most efficient CR driven MFA mechanism so far identified, the non-resonant streaming instability \cite{Bell04}, only allows a small fraction of SNRs to act as PeVatrons and only for a small fraction of their lifetime \cite{Schure13,Cristofari2020,Diesing2023TheRays}.} While observational evidence for the magnetic field amplification was indeed recorded, e.g. through X-ray observations of variable hot spots in the young SNR RX J1713.7-3946 \citep{2008ApJ...677L.105U}, compelling evidence for hadronic gamma-ray emission above few hundreds TeV in correspondence of an SNR is still lacking. 

The advent of LHAASO (Large High Altitude Air Shower Observatory), a new extensive air shower facility located in China, allowed us to explore the highest energy regime of gamma rays, beyond PeV. In their first catalog they list 90 sources, among which 43 overcome 100 TeV, but no firm association is claimed with SNRs. It is impressive to note that a few SNRs seem to be coincident with LHAASO sources, but this concerns only old objects, where the PeVatron phase is believed to be over in the standard paradigm. Alongside, no young SNR emerges in the PeV Sky. A detailed data analysis of the young ($\sim$ 300 year old) SNR Cassiopeia~A \citep{2024ApJ...961L..43C} revealed no significant emission beyond 10 TeV, re-opening a debate regarding the contribution of this type of sources to the highest energy CRs. 

Challenges to the detection of \red{UHE} radiation in correspondence of SNRs could also derive in principle from the difficulty of confining high-energy particles inside the remnants. Indeed, it is expected that, as the velocity of the shock decreases, also the magnetic turbulence responsible for confining the accelerated particles declines, allowing the highest energy particles to leave the remnant very soon \citep{2020A&A...634A..59B,2021A&A...650A..62C}. \giada{In this case, UHE emission could be possibly detected from the surroundings of the remnant, provided that the target density for hadronic interactions is sufficiently high. }  
This hypothesis was recently tested by \cite{Mitchell2024} who scanned the UHE catalogs to find possible counterparts in molecular clouds illuminated by SNRs, however, due to the large uncertainties, the results were inconclusive. 
{Complementary information derived from low-energy observations could {be used to clarify the acceleration history in SNRs}. {Although models of particle acceleration at SNRs typically agree on the idea that higher energy particles are released earlier than low energy ones, the time and energy dependent escape of CRs from a SNR (and in general from any accelerator) is poorly understood. This problem is particularly challenging to address,both analytically and through numerical simulations, due to the variety of scales (from the particle Larmor radii or even smaller, to tens pc scales) and conditions (from the high turbulence level at the shock, to the much lower level of typical ISM conditions) {that are relevant for} the escaping particles. In addition, escaping CRs often excite plasma instabilities, thus making the escape process itself non-linear \cite{Malkov2013}.  All these aspects should be carefully taken into account, since the energy-dependent escape and propagation outside the source may produce quite different CR spectra close to the shock or in a MC located at a given distance, with a non-trivial dependence on the source age \citep{Gabici2007,Gabici2009Broad-bandRemnants,2012A&A...541A.153T}. 
\subsubsection{\giada{Promising SNRs as targets for MeV observations}}
To date, the \fermi~ 4th catalog of sources \cite{4fgl} \red{(4FGL)} counts 24 sources firmly identified with SNRs, 19 sources possibly associated with SNRs and additional 114 sources which are of ambiguous identification between an SNR and a PWN.
\red{For comparison, the radio‐based Green catalog now lists some 310 Galactic SNRs \cite{Green-cat-SNR-2025}.}
In the TeV sky, 8 SNRs have been firmly identified by the \hess Galactic plane survey and as many are identified as composite objects, namely SNRs that embed a PWN. The association of LHAASO sources with SNRs, instead, remains controversial: \giada{as anticipated, UHE emission is reported in correspondence of two middle-age SNRs, namely W51~C and IC~443 \citep{lhaaso_w51,lhaaso_ic433} by the LHAASO collaboration, but the interpretation is challenged by the superposition of multiple sources.   } 

\red{A wide range of morphologies has  emerged for SNRs from gamma-ray observations: for a portion of them a shell-like structure \giada{that nicely corresponds to the X-ray morphology} was unveiled at VHE, while other objects show spatially complex emission, sometimes extending beyond the SNR boundary and \ea{likely} tracing the interaction of escaping particles with surrounding MCs. 
One example of the first type is the case of 
RX~J1713-3946, \ea{a young ($<2000$ yr)} SNR observed by \hess \citep{HessRXJ}, while the \giada{middle-aged ($>10^4$ yr)} SNR W28 \cite{Fermi-W28-2010} is  a representative case of the second type \giada{: here the emission does not match the SNR shock, but rather coincides with peaks of external gas clouds} (see Fig. \ref{fig:snr_hess}) \citep{hess_w28}. \giada{The effect on the clouds is two-fold: the interaction of accelerated particles with the gas produces gamma-rays and contributes to increase the ionization rate of the region \citep{Vaupre_W28IonRate_2014,phan2020}}.
The spectral energy distributions (SEDs) detected from SNRs are also very different, with a tendency to be  harder for young objects, and steeper in middle-aged SNRs, as summarized in \cite{funk}.
} 

\begin{figure}
    \centering
    \includegraphics[width=1\linewidth]{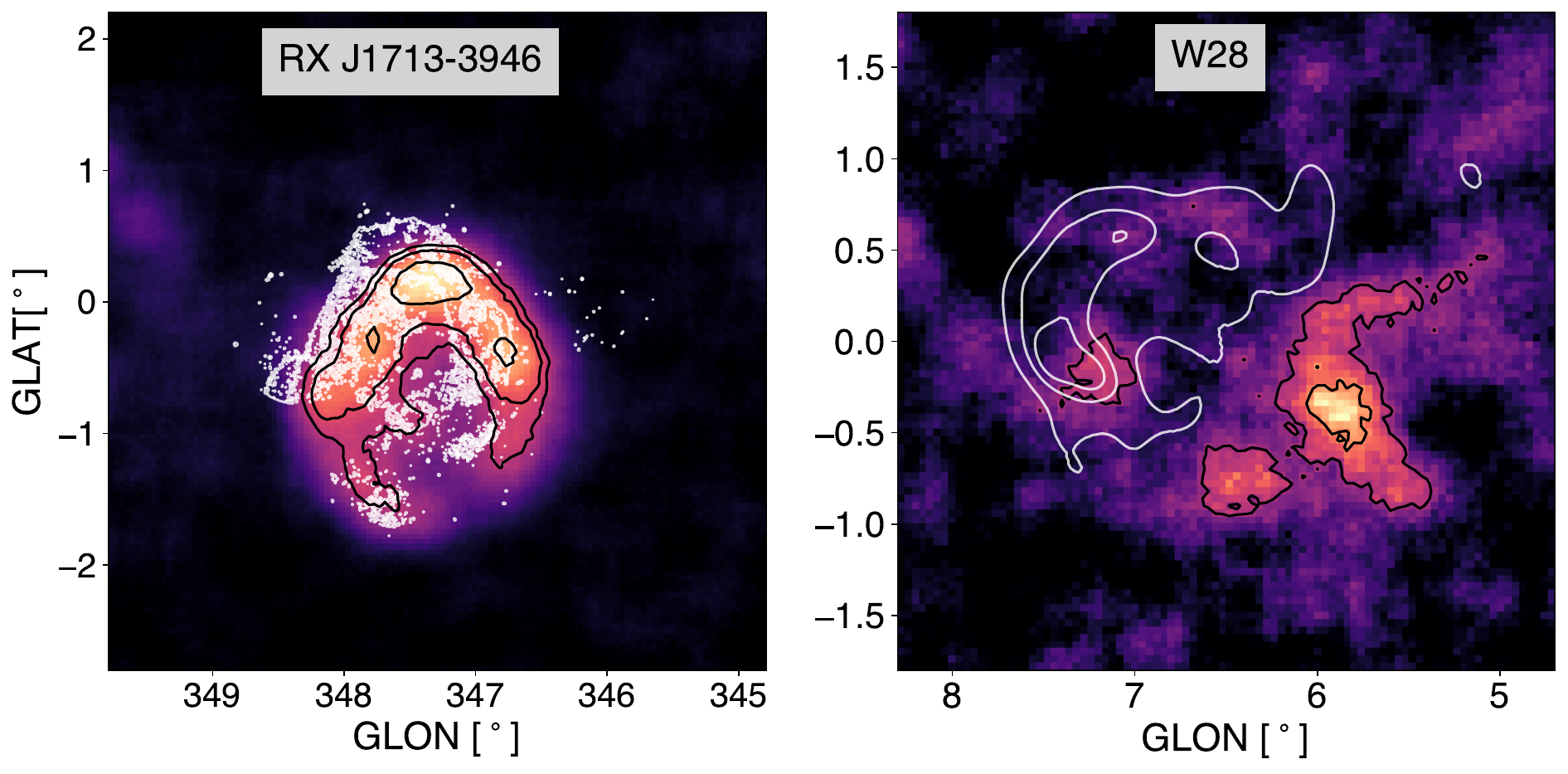}
    \caption{Supernova remnant detected by \hess \citep{hgps}: the map represents the integrated flux in the band 1-10 TeV. On the left the \giada{young shell-type} SNR  RX~J1713.7-3946 \giada{; black contours indicate the H.E.S.S. measurements, while the white contours indicate X-ray emission from CHANDRA/ACIS. On the right: the middle-age SNR W28; in this case the white contours represent the radio emission from the GLEAM survey at 70 MHz.}  }
    \label{fig:snr_hess}
\end{figure}
\red{This variety is not unexpected, as both electrons and ions produce gamma-ray emission, and both the spectrum and morphology of their respective contributions evolve differently \cite{2020A&A...634A..59B}. \giada{As far as leptonic emission is concerned,} Inverse Compton emission of multi-TeV electrons quickly fades \ea{with increasing distance from the acceleration site} because of the severe energy losses suffered by leptons. \giada{On the other hand, as far as hadronic emission is concerned, one important factor that affects both the morphology and the spectrum is that} the maximum energy to which the SNR shock can accelerate \ea{hadrons} is expected to decrease with time, due to the slowing down of the shock and to the declining ability of the accelerated particles to efficiently drive turbulence. \giada{In terms of spectrum, this implies that} the cumulative particle yield tends to be softer above the maximum energy presently achievable \cite{2019MNRAS.490.4317C,2020A&A...634A..59B}. In terms of morphology, the spatial distribution of particles illuminating nearby MCs is \ea{determined} by the time-dependent escape of CRs from the shock and the evolution of their maximum energy, combined with the subsequent propagation from the shock to the cloud. All these ingredients are still poorly constrained. This picture is further complicated by the structure of the wind bubble in which core-collapse supernovae explode \cite{2024A&A...689A...9D}. \giada{Despite these expected qualitative morphological differences, the identification of the nature of the emission from a number of SNRs (including e.g. RX~J1713-3946 shown in Fig. \ref{fig:snr_hess}) is still debated \citep[e.g][]{Cellirxj} and can be investigated with MeV observations of primary and secondary emission as we will describe in  more detail in Sec. \ref{sec:signatures}.}
}

In Fig.~\ref{fig:snr_cat} we show the SED of the SNRs reported in the 4FGL catalog, together with the sensitivity of upcoming MeV detectors \citep{Amego2022,BergeICRC}. For many of these sources dedicated analyses are available, where specific regions are better optimized. However, the data from the catalog can be used for a rough estimate of the flux level expected from these sources in the MeV-GeV band. Even  with an exposure of a few years, the planned MeV detectors will be able to characterize the low-energy part of most of the sample. \giada{As already stressed,} this will be important to characterize the acceleration and the emission mechanisms. 

\begin{figure}
    \centering
    \includegraphics[width=0.9\linewidth]{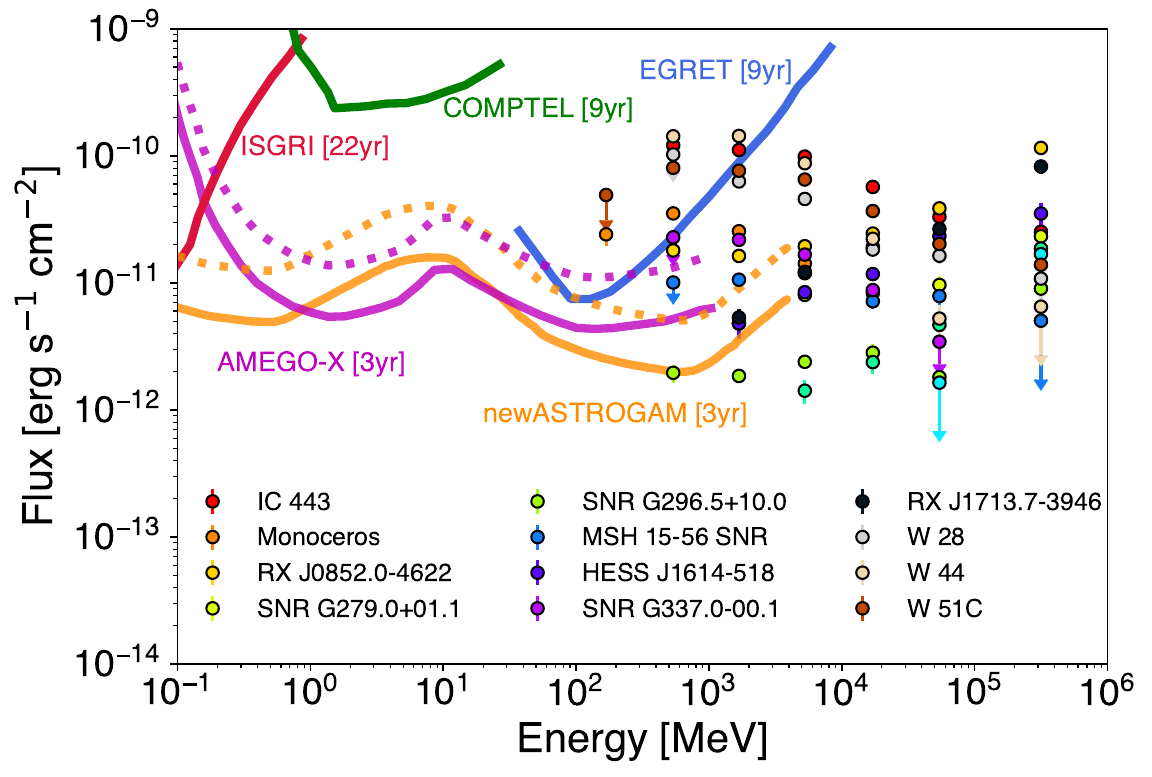}
    \caption{Spectral energy distribution as derived from \fermi~observations towards \giada{a sub-sample of} supernova remnants \giada{of the 4FGL catalog}\citep{4fgl}. \red{We report the flux only for those energy bins where the test-statistic value is larger than 2: flux-}points with 2$<$TS$<$4 are reported as upper limits\red{, while flux-points with TS $>$4 are reported with their relative 1 $\sigma$ uncertainty.} The sensitivity curves for future MeV detectors AMEGO-X (magenta) \citep{Amego2022} and newASTROGAM \citep{BergeICRC} (orange) are also plotted \giada{together with the sensitivity of the past MeV detectors COMPTEL, INTEGRAL, and EGRET, for different exposure times \citep{BergeICRC}}. \red{As most SNRs detected by Fermi-LAT are extended, we report the sensitivity curves also for the case of extended sources (dotted curves), considering the average extension of catalogued SNRs, namely 0.7$^{\circ}$.}}
    \label{fig:snr_cat}
\end{figure}
 
\subsection{Stars and Stellar Clusters}\label{subsec:SC}
{Clusters of young massive stars (YMSCs) might meet, under certain conditions, the requirements for being Galactic PeVatrons \citep{AharonianYMSC,Morlino_2021,blasi_morlino2023}. 
\red{The advantage of YMSCs over SNRs is twofold: on one hand the shock velocity remains roughly constant over a timescale (1-10 Myr), allowing for a longer confinement of particles; on the other hand, the bubble which is excavated by the winds is hot and turbulent, therefore it is likely characterized by a reduced diffusion coefficient compared to the interstellar one, that allows the acceleration time to be reduced. As discussed in \ref{sec:acc_trans} acceleration in SNRs is challenged both by the decrease of the shock velocity over time, and by the fact that the shock expands through the interstellar medium and therefore  magnetic field amplification is required \citep[e.g.][]{blasirev} in order to reach PeV energies. Following the derivation of \citep{Morlino_2021}, it is easy to see that if a substantial fraction of the wind power of a YMSCs is converted into MHD turbulence, a powerful cluster ($L_w\gtrsim10^{37}$~erg~s$^{-1}$) can reach PeV energies (see Eqs. \ref{eq:EmaxB},\ref{eq:EmaxKra},\ref{eq:EmaxK41}). } 
The power released by the winds of massive stars in a cluster can be comparable to the kinetic luminosity of supernova explosions in the Galaxy \citep{Seo2018}, and the conditions for efficient particle acceleration might be met in the turbulent environment that the interaction of multiple stellar winds creates.} 
The powerful winds from OB-type stars in YMSCs are able to deeply shape the surrounding environment by inflating large cavities in the interstellar medium (ISM), commonly referred to as SC \textquotedblleft bubbles". For compact YMSCs\footnote{We define compact YMSCs as those clusters where the winds from massive stars can combine together to create a collective cluster wind. Although there is no absolute criterion to determine if a cluster is compact, a good approximation to define a compact cluster is that the radius of the collective wind termination shock is larger than the cluster half-mass radius \citep{Morlino_2021}.}, the structure and evolution of these wind-blown bubbles are similar to that of the bubbles that develop around isolated massive stars \citep{Weaver_1977}. The bubble structure is divided into 3 regions (see also Figure \ref{fig:WindBubble}):
\begin{itemize}
    \item An inner part, occupied by the \red{ supersonic}   cold cluster wind.
    \item A low-density cavity filled with the hot shocked wind material.
    \item A dense shell of swept-up material created by the expansion of the hot cavity.
\end{itemize}

The cold inner region and the hot cavity are separated by the wind termination shock (TS), while the hot cavity and the swept-up shell are connected by a contact discontinuity. The forward shock (FS) is what separates the swept-up shell from the unperturbed ISM. Due to intense radiative losses in the high density environment, the swept-up shell collapses into a tiny layer after a few kyrs \citep{Castor_1975}, so that the radius of the contact discontinuity ($R_{cd}$) and the radius of the FS ($R_{fs}$) almost coincide ($R_{fs} \simeq R_{cd}$). As a consequence, the structure of the bubble can be simplified into a two zone model by neglecting the swept-up shell region. Figure~\ref{fig:WindBubble} shows a sketch of this simplified two-zone model of the wind-blown bubble together with the image of the wind-blown bubble created by the YMSC NGC~2244.

{Assuming that the expansion of the hot cavity is adiabatic}\footnote{Radiative losses in the hot cavity are expected to become dominant only after $\sim 10$ Myr \citep{Silich_2013}, unless turbulent mixing of cold ISM material, induced by the fragmentation of the thin swept-up shell, cools down the hot cavity to temperatures where radiation cooling becomes efficient \citep{Lancaster_2021}.},  the size of the TS ($R_{\rm ts}$) and the FS radius can be calculated as \citep{Weaver_1977}:
\begin{equation}
\label{eq:Rts}
R_{\rm ts} = \sqrt{\frac{(3850 \pi)^{2/5}}{28 \pi}} \, \dot{M}^{1/2} v_{\rm w}^{1/2} L_{\rm w}^{-1/5} \rho_0^{-3/10} t_{\rm age}^{2/5} 
,\end{equation}
\begin{equation} \label{eq:Rfs}
  R_{\rm fs} =\left( \frac{125}{154 \pi} \right)^{1/5} 
  L_{\rm w}^{1/5} \rho_0^{-1/5} t_{\rm age}^{3/5} 
,\end{equation}
where $\rho_0$ is the density of the local ambient medium in which the bubble expands, $t_{\rm age}$ is the age of the YMSC, $L_{\rm w}$ is the wind kinetic luminosity, and $\dot{M}$ and $v_{\rm w}$ are the mass-loss rate and the velocity of the cluster wind, respectively.

\begin{figure}
\centering
\includegraphics[width=0.48 \textwidth]{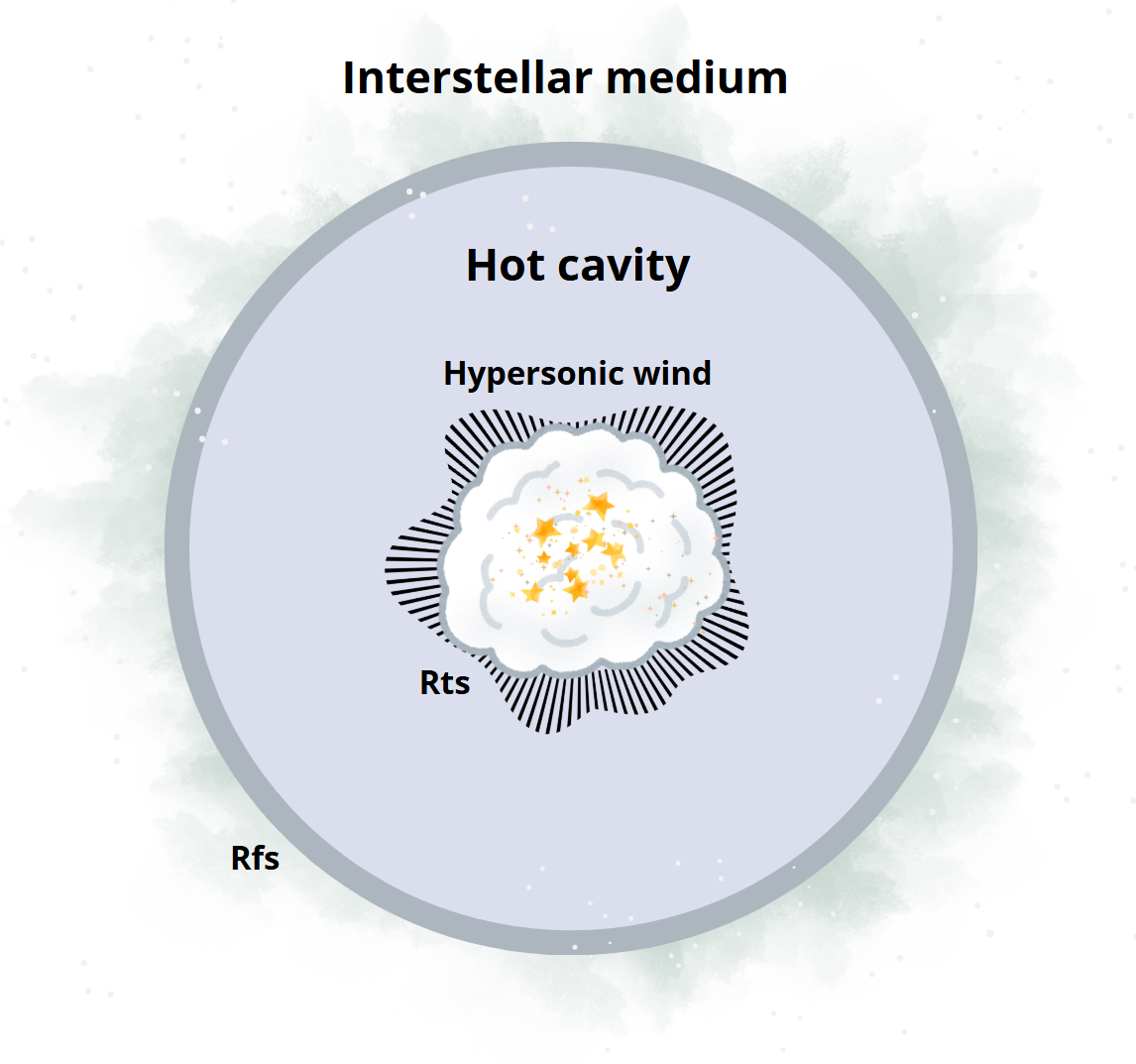}\includegraphics[width=0.48 \textwidth]{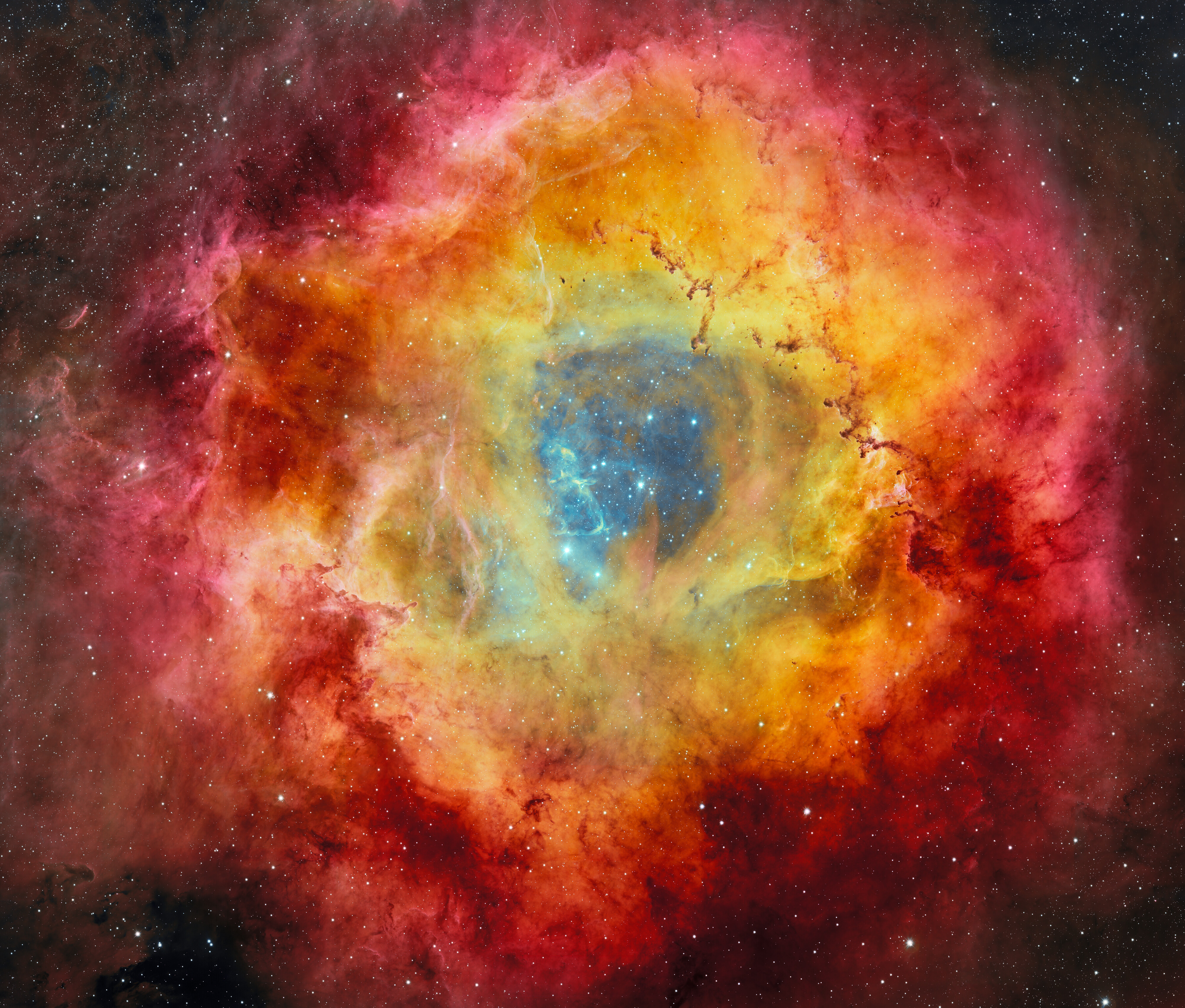}
     
\caption{Left panel: Two zone scheme of the structure of a wind-blown bubble. Right panel: Example of a wind-blown bubble from the young massive star cluster NGC~2244. The image is taken with the Dark Energy Camera, mounted on the U.S. National Science Foundation Víctor M. Blanco 4-meter Telescope at Cerro Tololo Inter-American Observatory. {[Credit: CTIO/NOIRLab/DOE/NSF/AURA. Image Processing: T.A. Rector (University of Alaska Anchorage/NSF NOIRLab), D. de Martin and M. Zamani (NSF NOIRLab)]}}
    \label{fig:WindBubble}
\end{figure}

In this specific systems, particle acceleration can occur at the strong collective cluster wind TS. This scenario has been investigated in a recent paper by \cite{Morlino_2021}, who provided a semi-analytical solution for the CR distribution in the wind blown bubble. We refer to that for a comprehensive discussion on how to obtain the CR distribution inside YMSCs. 

The maximum momentum reached by accelerated particles can be determined by equating the diffusion length, upstream of the wind TS (\red{i.e. in the supersonic free wind}) to the radius of the TS, namely:
\begin{equation} \label{eq:PmaxCond}
    \frac{ D_{\rm w}(p_{\max})}{v_w} = R_{\rm ts}\, .
\end{equation}

{The diffusion coefficient depends on the properties of the ambient turbulence, which are largely unknown, both in terms of strength and spectrum.
Therefore, the best one can do is to consider several widely used models for astrophysical turbulence, and the ensuing diffusion coefficients, namely the Kolmogorov ($D_{\rm Kol}$), Kraichnan ($D_{\rm Kra}$) and Bohm ($D_{\rm B}$) diffusion coefficients:}

\begin{subequations} \label{eq:DiffCoeff}
\begin{equation}
D_{\rm Kol}=\frac{1}{3} v_p r_{\rm L}^{1/3} L_{\rm inj}^{2/3}
,\end{equation}
\begin{equation}
D_{\rm Kra}=\frac{1}{3} v_p r_{\rm L}^{1/2} L_{\rm inj}^{1/2}
,\end{equation}
\begin{equation}
D_{\rm B}=\frac{1}{3} v_{\rm p} r_{\rm L}\,,
\end{equation}
\end{subequations}
where $L_{\rm inj}$ is the length scale at which the magnetic turbulence is injected\footnote{In YMSCs this is often assumed to be of the order of the average distance between stars ($\sim 1$~pc).}, and $v_p$ and $r_{\rm L}$ are the particle velocity and Larmor radius respectively. 

{The different prescriptions result in different estimates of the maximum energy achievable in these environments ($E_{\max}=p_{\max}\,c$):} 

\begin{align}
 E_{\max}^{\rm Kol} &\simeq 1.2 \left(\frac{\eta_{\rm B}}{0.1}\right)^{1/2} \left( \frac{\dot{M}}{10^{-4} \rm M_\odot yr^{-1}}\right )^{-3/4} \left( \frac{L_{\rm w}}{10^{39} \rm \, erg\,s^{-1}} \right )^{37/20} \nonumber \\ 
   & \left( \frac{\rho_0}{20\ m_p \, \rm \ cm^{-3}} \right )^{-3/5} \left(\frac{t_{\rm age}}{3 \rm \ Myr}\right )^{4/5} \left( \frac{L_{\rm inj}}{2 \rm \ pc} \right )^{-2} \rm PeV \label{eq:EmaxK41} \,,    
\end{align}
\begin{align}
 E_{\max}^{\rm Kra} &\simeq 2.84 \left(\frac{\eta_{\rm B}}{0.1}\right)^{1/2} \left( \frac{\dot{M}}{10^{-4} \rm M_\odot yr^{-1}}\right )^{-5/10} \left( \frac{L_{\rm w}}{10^{39} \rm \ erg \, s^{-1}} \right )^{13/10} \nonumber\\   
   & \left( \frac{\rho_0}{ 20\, m_p \rm \ cm^{-3}} \right )^{-3/10} \left(\frac{t_{\rm age}}{3 \rm \ Myr}\right )^{2/5} \left( \frac{L_{\rm inj}}{2 \rm \ pc} \right )^{-1} \rm PeV \label{eq:EmaxKra}   \,,
\end{align} 
\begin{align}
 E_{\max}^{\rm B} &\simeq 10 \left(\frac{\eta_{\rm B}}{0.1}\right)^{1/2} \left( \frac{\dot{M}}{10^{-4} \rm M_\odot yr^{-1}}\right )^{-1/4} \left( \frac{L_{\rm w}}{10^{39} \rm \ erg\, s^{-1}} \right )^{3/4} \rm PeV \label{eq:EmaxB}   \,.
\end{align}
where $\eta_{\rm B}$ is the fraction of the wind kinetic power dissipated into magnetic turbulence, usually assumed to be of the order of 10\%.
{We see that depending on the mass-loss rate and wind luminosity of the system, very high energies can in principle be achieved. These quantities depend only on the mass and age of the systems.} {These parameters were recently derived \cite{Celli_2024}} 
{for a sample of optically identified star clusters \cite{Cantat-Gaudin} located within a few kpc from us: these can be seen in Figure \ref{fig:SC}. Comparing the estimated mass-loss rates and wind luminosities with the prediction for the maximum energies presented in Eqs. \ref{eq:EmaxK41},\ref{eq:EmaxKra},\ref{eq:EmaxB}, one finds that only a few SCs of this sample are able, even in principle, to reach PeV energies. {In Table~\ref{tab:Emax_sc} we list a few interesting candidates where, depending on the type of turbulence, one would expect high or very-high energy emission. } This can also be seen from Figure \ref{fig:SC_emax}, where a theoretical prediction for the maximum energy achievable in the case of Kraichnan diffusion is shown for the sample of clusters selected by \cite{Celli_2024}. 

\red{
From the point of view of MeV observations, SCs can be interesting in two ways. On the one hand, objects that are bright gamma-ray emitters at GeV energies \cite{Peron_2024} are expected to remain bright also at slightly lower photon energies, so that currently detected SCs are also promising targets for MeV observations. On the other hand, under the assumptions presented above, only a small fraction of clusters are capable of accelerating particles up to the highest energies. In contrast, a large number of low-mass clusters are not expected to exceed GeV energies and are therefore likely to emit predominantly in the MeV domain. }
Considering that the sample identified in \cite{Cantat-Gaudin} represents less than 1/3 of all the clusters hosted in the Milky Way, and that in the Galaxy low-mass SCs are more numerous than massive ones, it is likely that many such sources will emerge in the MeV domain.

\begin{figure}
    \centering
    \includegraphics[width=1\linewidth]{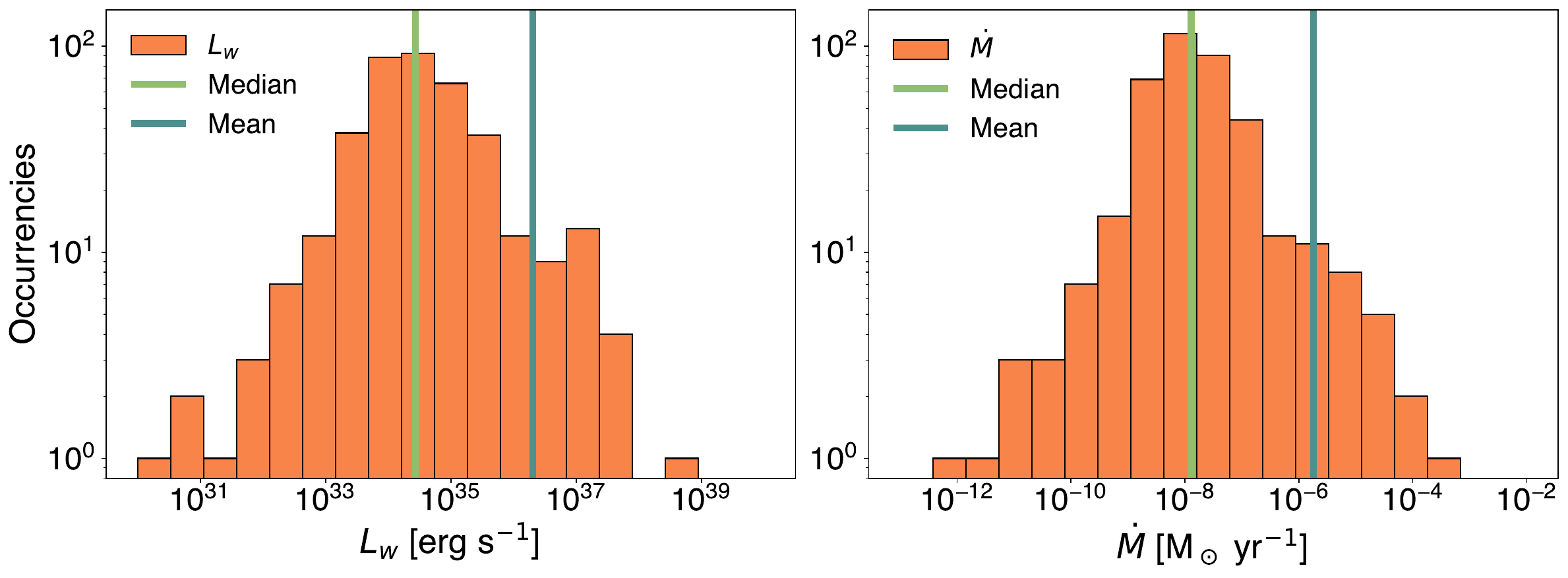}
    \caption{Physical properties of the sample of star clusters listed in \cite{Celli_2024}.{The left panel shows the distribution of the calculated wind luminosity, while the right side shows the mass-loss rate.} In both cases the mean and the median are indicated as a vertical line in the distributions.  }
    \label{fig:SC}
\end{figure}{}

\begin{figure}
    \centering
    \includegraphics[width=1\linewidth]{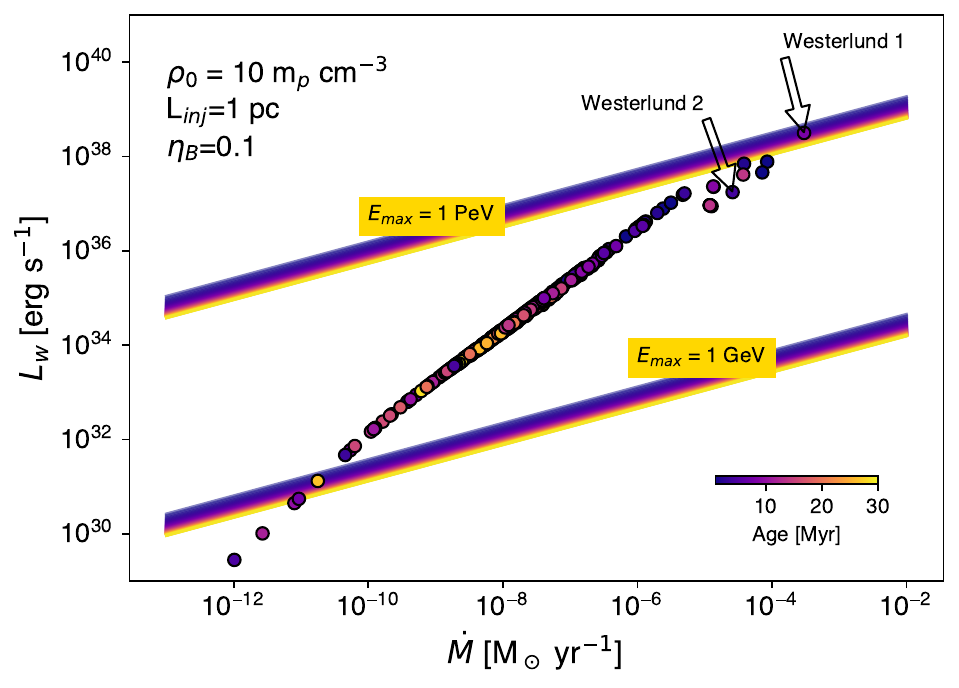}
    \caption{Maximum achievable energy in star clusters depending on their age, wind luminosity and mass loss-rate, calculated in the case of \red{Kraichnan} diffusion as described in Eq. \ref{eq:EmaxKra}, assuming $\rho_0 = 10~m_p~cm^{-3}$, $L_{inj}=1~$pc $\eta_B=0.1$. The sets of lines indicate, for a given a value of $\dot{M}$ and a given age of the cluster (see the colorbar), the minimum  $L_{w}$ required to reach GeV and PeV energies (see labels in the yellow boxes).The points show the combination of parameters ($L_w$, $\dot{M}$, and age) obtained for the set of cluster parameters listed in \citep{Celli_2024}, with ages taken from the compilation of \citep{Cantat-Gaudin}. A cluster can reach the energy specific of each set of lines (1 PeV or 1 GeV) only if its corresponding point overcomes the line of the set that corresponds to the age of the cluster. With these assumptions, Westerlund~1 is the only cluster in the sample able to reach 1 PeV.}
    \label{fig:SC_emax}
\end{figure}{}

\begin{table}[]
\begin{tabular}{cccccccc}
Name         & \begin{tabular}[c]{@{}c@{}}Mass\\ {[}M$_\odot${]}\end{tabular} & \begin{tabular}[c]{@{}c@{}}Age\\ {[}Myr{]}\end{tabular} & \begin{tabular}[c]{@{}c@{}}L$_{\rm w}$\\ {[}erg s$^-1${]}\end{tabular} & \begin{tabular}[c]{@{}c@{}}$\dot{M}$\\ {[}M$_\odot$ yr$^{-1}${]}\end{tabular} & \begin{tabular}[c]{@{}c@{}}E$_{\max}^{\rm Kol}$\\ {[}TeV{]}\end{tabular} & \begin{tabular}[c]{@{}c@{}}E$_{\max}^{\rm Kra}$\\ {[}TeV{]}\end{tabular} & \begin{tabular}[c]{@{}c@{}}E$_{\max}^{\rm B}$\\ {[}TeV{]}\end{tabular} \\
 \midrule
NGC 2244     & 2096    & 12.6       & $2.7 \times 10^{35}$      & $1.2 \times 10^{-7}$   & 1       & 10       & 110   \\
NGC 3603     & 4808    & 1& $4.6 \times 10^{37}$      & $7.2 \times 10^{-5}$   & 13      & 100      & 1100  \\
NGC 6357     & 1387    & 1& $4.2 \times 10^{36}$      & $1.3 \times 10^{-6}$   & 3       & 30     & 480   \\
NGC 6611     & 1697    & 2.1        & $6.3 \times 10^{36}$      & $2 \times 10^{-6}$     & 5       & 40      & 590   \\
Trumpler 16  & 1106    & 13.5       & $9.2 \times 10^{36}$      & $1.2 \times 10^{-5}$   & 20      & 80     & 500   \\
Westerlund 1 & $2.2 \times 10^{4}$    & 7.9        & $3.1 \times 10^{38}$      & $3 \times 10^{-4}$     & 800     & 1300    & 3100  \\
Westerlund 2 & 2172    & 4& $1.8 \times 10^{37}$      & $2.6 \times 10^{-5}$   & 15      & 80      & 670 \\  

\end{tabular}
\caption{Theoretical maximum energy of accelerated particles in some Galactic YMSCs. {External density and magnetic energy fraction are assumed as discussed in the text.} 
The clusters parameters are obtained from \cite{Celli_2024}.}
\label{tab:Emax_sc}
\end{table}

It is worth noting that, although the model for acceleration presented in \citep{Morlino_2021} was developed for compact clusters, it can also be applied to isolated massive stars. In this case, particle acceleration will occur at the TS of the stellar wind. Table~\ref{tab:Emax_star} shows the maximum particle energies achievable by single massive stars, for different values of stellar masses. To determine these values of $E_{\max}$, we assumed $t_{\rm age}=1$~Myr, $\eta_{\rm B}=0.1$, $\rho_0=1~m_p \rm \, cm^{-3}$, and that magnetic turbulence is injected with a typical length scale equal to a fraction of the TS radius ($L_{\rm inj}=0.1\, R_{\rm ts}$), while the stellar wind power and mass loss rates have been calculated using the empirical approach described in \cite{Menchiari_2024b}. One can notice that the maximum energies are higher than 1~TeV, implying that also single massive stars can be significant sources of sub-GeV particles. 

Figure \ref{fig:emax_bubble} summarizes the maximum energies achieavable in different wind-powered scenarios \giada{considering the most effective diffusion type, namely a Bohm-like diffusion coefficient (see Eq. 17)}, and varying only the typical mass-loss-rate and the typical wind velocity. In order to reach very- and ultra-high energies, extra fast winds and sizeable mass losses are needed. Aside from star clusters, when looking for alternative PeVatrons, it is worth noticing that extreme conditions in terms of outflow velocity and mass loss rate are realized, for example, in ultra-luminous X-ray sources (ULXs), where an accreting stellar-mass compact object is believed to launch a wind, similar to a star cluster wind, but with a much larger velocity and ejected mass \citep{PerettiULX}. On the other hand, if ULXs \giada{(see discussion in \citep{AmatoISSI})} and SCs may contribute to the highest energies, it is possible that the GeV and sub-GeV domain is dominated by individual massive stars \giada{whose winds 
\ea{are not powerful enough to provide particle acceleration up to energies exceeding} a few GeV even in the most promising conditions (see Table \ref{tab:Emax_star})}.  

\begin{table}[]
\begin{tabular}{cccccc}
\begin{tabular}[c]
{@{}c@{}} Star Mass\\ {[}M$_\odot${]}\end{tabular} & \begin{tabular}[c]{@{}c@{}}L$_{\rm w}$\\ {[}erg s$^-1${]}\end{tabular} & \begin{tabular}[c]{@{}c@{}}$\dot{M}$\\ {[}M$_\odot$ yr$^{-1}${]}\end{tabular} & \begin{tabular}[c]{@{}c@{}}E$_{\max}^{\rm Kol}$\\ {[}TeV{]}\end{tabular} & \begin{tabular}[c]{@{}c@{}}E$_{\max}^{\rm Kra}$\\ {[}TeV{]}\end{tabular} & \begin{tabular}[c]{@{}c@{}}E$_{\max}^{\rm B}$\\ {[}TeV{]}\end{tabular} \\
 \midrule
10      & $2.3 \times 10^{34}$  & $1.2 \times 10^{-8}$         & 1            & 5          & 30         \\
25      & $1.5 \times 10^{36}$  & $5.7 \times 10^{-7}$         & 20           & 60           & 290        \\
50      & $1.9 \times 10^{37}$  & $5.8 \times 10^{-6}$         & 80           & 233          & 1000       \\
75      & $5.7 \times 10^{37}$  & $1.6 \times 10^{-5}$         & 150         & 440          & 1800       \\
100     & $1.2 \times 10^{38}$  & $3.1 \times 10^{-5}$         & 250          & 690          & 2800       \\
\end{tabular}
\caption{Theoretical maximum energy of accelerated particles in massive stars. The stellar mass loss rate and wind power are calculated using the approach described in \cite{Menchiari_2024b}. }
\label{tab:Emax_star}
\end{table}

\begin{figure}
    \centering
    \includegraphics[width=0.9\linewidth]{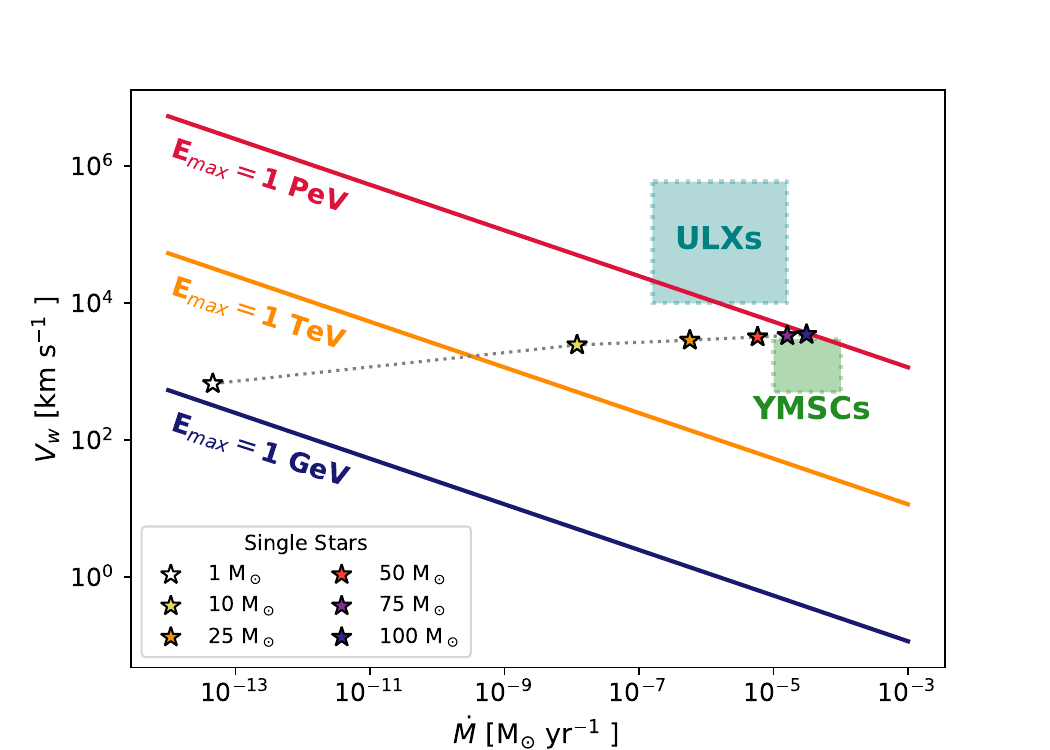}
    \caption{Typical maximum energies reached by different stellar ecosystems  (ULXs, SCs and single stars) \giada{considering a Bohm confinement}, depending on their wind luminosity and mass-loss rate. Typical values for these quantities are indicated. [Figure adapted from \citep{PerettiULX}] }
    \label{fig:emax_bubble}
\end{figure}

\subsubsection{\red{Gamma-ray detections of star clusters}}
The detection, {obtained by H.E.S.S.}, of very-high-energy photons from the {Galactic} YMSCs Westerlund~1 and Westerlund~2 \citep{2007A&A...467.1075A,2022A&A...666A.124A} {and from the clusters 30~Doradus~C, and R136 in the Large Magellanic cloud \citep{LMC_HESS_2015,2024ApJ...970L..21A} {strengthen the case, already made by {\it Fermi}, with the detection in the GeV band of several such sources (like NGC~3603 \citep{Sun3603,SahaNGC3603,Rocamora3603}, NGC~6618 \citep{Liu_M17}, and, RCW~38 \citep{PeronVela,Pandey2024,Ge2024}), that star clusters are effective at accelerating particles well above TeV energies.}
{In addition, in the LHAASO first catalog of ultra high energy sources \citep{Cao2024},} {the highest energy source, and the one for which \red{a} hadronic origin of the emission seems most convincing \cite{deona22}, is spatially coincident} with the Cygnus-X star-forming region, \citep{CygnusLHAASO}, already detected by Fermi-LAT \citep{Cygnus_Fermi,CygnusFermi_new}, {ARGO \citep{2014ApJ...790..152B}} and HAWC \citep{Cygnus_Hawc}, and containing the massive star association Cygnus OB2. The identification of the origin of the highest-energy photons, however, is still under debate \citep{Vieu_cygnus} as they may be contributed partly by the micro-quasar Cygnus X-3, at the center of the emitting region, but located much farther (at $\sim$7 kpc) than the star-forming region. Interestingly, the kinetic energy enclosed in the winds of the Cygnus OB2 stellar association has enough power to support the emission, but it is not straightforward to fit the multi-wavelength SED, as discussed in \cite{Menchiari_2024,blasi_morlino2023}.    }

The strong correlation found by \cite{Peron_2024} between H\textsc{II} regions produced by massive stellar clusters of ages below $\sim$~2~Myr (i.e. with no {supernova events expected to have occurred}) and unidentified Fermi-LAT sources indicates that {star clusters are generically able to accelerate particles} at least {up to the} GeV range, suggesting that these should contribute to some extent to the bulk of Galactic CRs.} 

The most massive stars begin to explode in supernovae after approximately 3 Myrs. In more evolved systems, multiple interacting winds of young massive stars and core collapse supernovae are {both} able to provide a large output of mechanical power, up to 10$^{39}$ \ergs, part of which can be converted into accelerated CR particles by shocks and turbulent plasma motions. Colliding-wind binaries are also possible CR accelerators that could be found in  massive star clusters and associations \citep[see e.g][]{1983SSRv...36..173C,2014A&ARv..22...77B,2019NatAs...3..561A,2023arXiv230106505G,2024NatAs...8..530P}, making the interpretation of the emission in these cases even more complicated. 
{This is where multi-wavelength observations, especially in the X-ray domain come into play.} {A {neat} example is provided by the} discovery by \cite{2007A&A...467.1075A} of the source of very high energy emission HESS~J1023-575 in the vicinity of the star forming region RCW~49 hosting the young open cluster Westerlund~2, which contains a very massive Wolf Rayet binary system, WR 20a. This discovery raised questions on the viable scenarios for the origin of the detected VHE emission. The authors of \cite{2007A&A...467.1075A} pointed out that the observed angular extension of HESS~J1023-575 {(0.18 $\pm$ 0.02)$^{\circ}$} disfavors {scenarios in which the emission is associated} with \red{a very} compact source like e.g. a colliding wind binary. However, the size of the acceleration site of VHE CRs can be smaller than the observed extension of HESS~J1023-575. 
The spatial resolution of IACTs is of the order of 0.1$^\circ$, {which corresponds to about 10~pc in terms of localization accuracy of a source at kpc distance.}
On the other hand, the size of the core, where most of the massive stars are localized in compact clusters, is just a few pc at most. Therefore, to establish or constrain the site of particle acceleration and radiation in clusters, as well as particle acceleration mechanisms, it is crucial to turn to the much higher (arcsecond) resolution of X-ray observations \citep[see e.g.][]{Kav20}. 
\chandra observations of the stellar cluster Westerlund~2 unveiled the presence of diffuse X-ray emission spreading over a region of several parsec. {The emission is interpreted as non-thermal, due to the large temperatures that would be needed otherwise to explain the observations in a thin thermal plasma scenario \citep{Townsley2005,Tow18}. Given the high radiation energy density in the region, inverse Compton scattering of stellar photons by mildly relativistic electrons was suggested as the origin of the non-thermal component of this diffuse halo \citep[see also][]{Wd2_Bedanarek97}.}}
{Similar conclusions were obtained for Westerlund~1, that was observed with \chandra by \citep{Mun06}, who, in addition to numerous point-like stellar sources, detected also a halo of diffuse X-ray emission, extending beyond the compact cluster core, but with an unclear origin. Recent attempts at detecting this halo towards Westerlund~1 were made with eRosita, however only unconstraining upper limits where derived \citep{Haubner2025EROSITA1}. }  

Recently, MHD modeling of the plasma flow produced by colliding winds in stellar clusters \cite{2022MNRAS.517.2818B,2024MNRAS.527.3749B} suggested the presence of magnetic fields of strength exceeding 100 $\mu$G. Therefore, a synchrotron origin of keV X-ray photons in the close vicinity of the cluster should not be disregarded, if multi-TeV CRs are accelerated in the cluster core. A magnetic field with filamentary structure and intensity  around 100 $\mu$G and higher, is seen in simulated clusters with a population resembling the one of \red{Westerlund~1}. A map of the magnetic field strength is shown in the right panel of Fig. \ref{fig:wd2-art}.

{The interpretation of non-thermal X-ray emission from the central regions of star clusters is always ambiguous, between a synchrotron and Inverse Compton origin, given the abundance of optical photons produced by luminous massive stars.}
A way to distinguish between the synchrotron and inverse Compton models is to search for MeV regime photons. Indeed the spectrum of IC emission would most likely extend to MeV photon energies (which is well before the transition to the Klein-Nishina regime) contrary to the synchrotron radiation of the cluster which is not expected to go beyond 100 keV. MeV detectors should be able to detect fluxes about 10$^{-14}$ \fluxcgs to address the issue in Westerlund~2 while for Westerlund~1 the sensitivity requirement is less stringent. 

\begin{figure*}
\centering
\includegraphics[width=160pt]{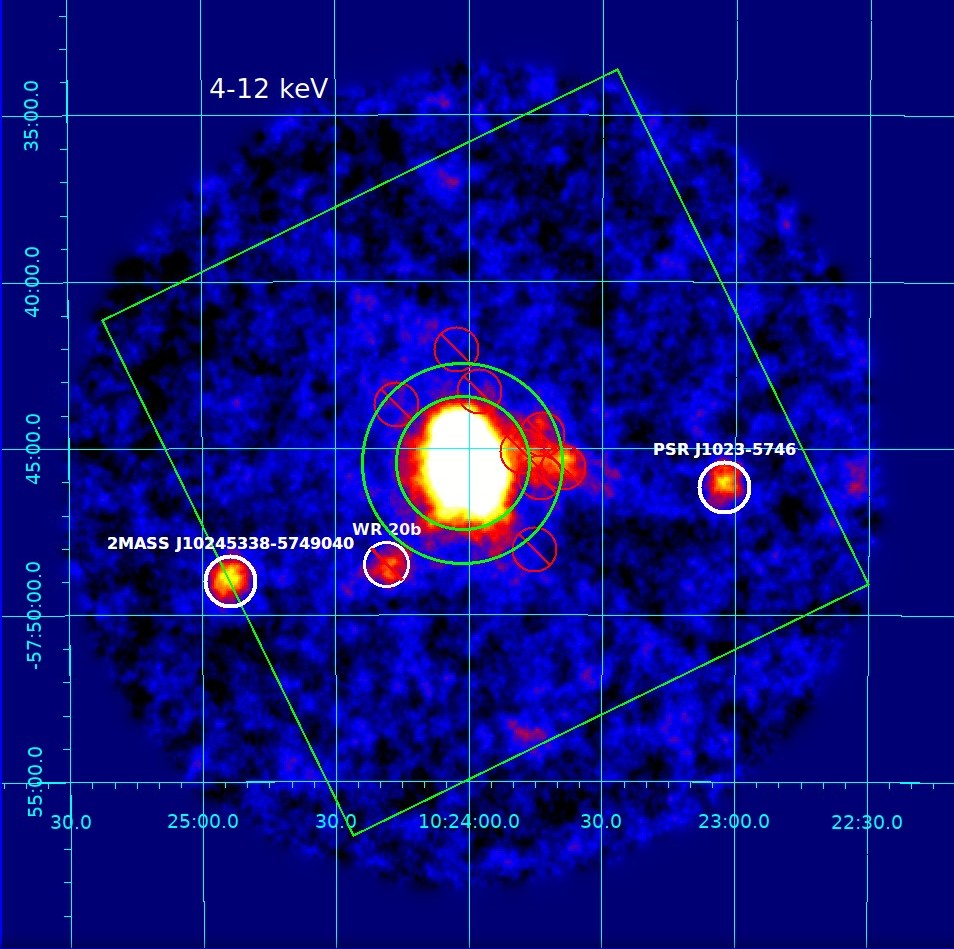}
\includegraphics[width=190pt]{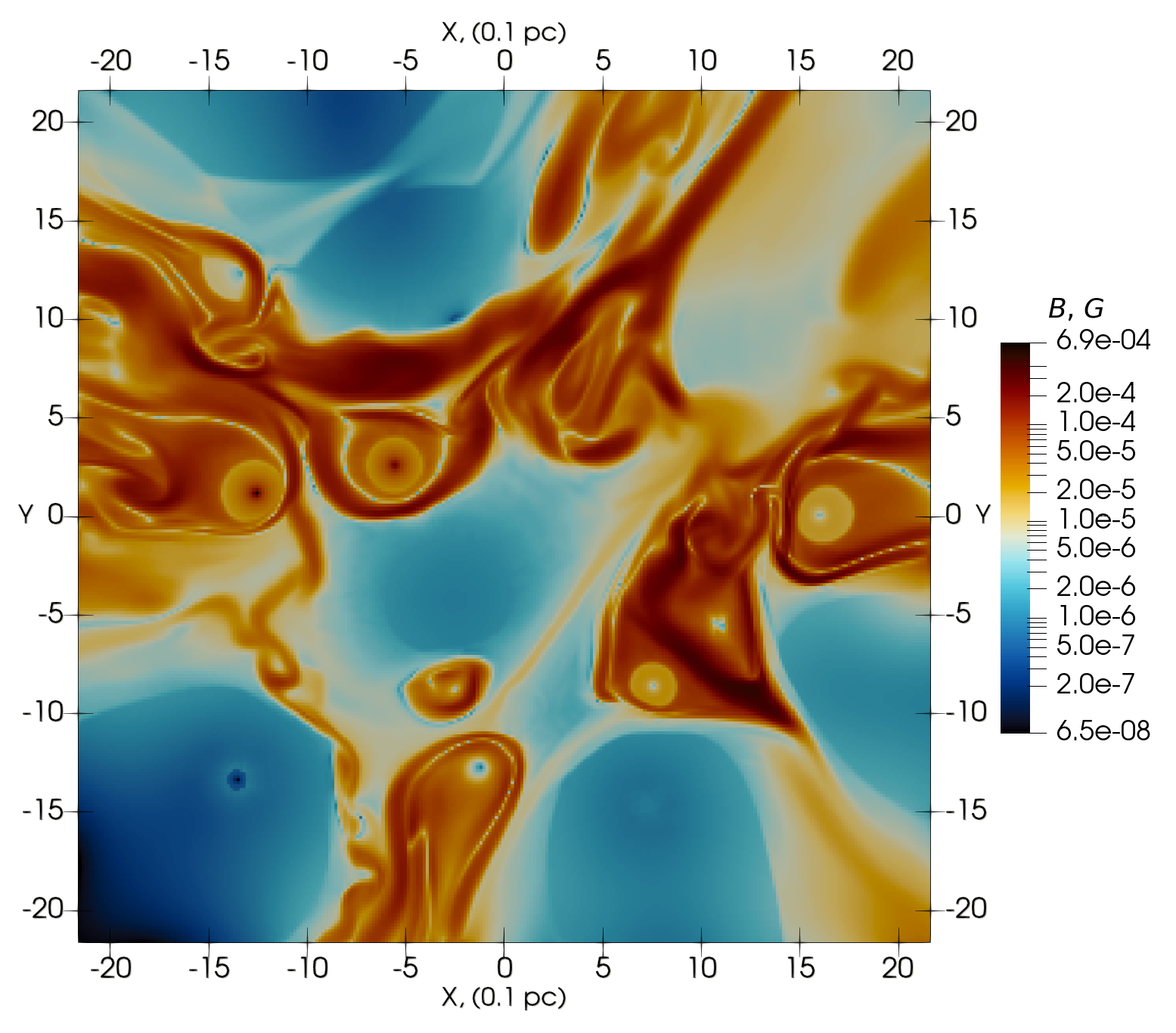}	
    \caption{Left panel: map of Westerlund 2 in the energy range 4-12 keV, as observed by the ART-XC telescope aboard SRG observatory. The annulus region used for the search of non-thermal X-ray emission is shown in green.  The FOV of \chandra\ used for a joint spectral analysis with ART-XC is shown as the box. The point sources excluded from the cluster field in the spectral analysis of the diffuse emission presented in \citep{2023MNRAS.525.1553B} are shown in red circles.
 Wolf-Rayet star WR20b, PSR J1023-5746, and the young stellar object candidate 2MASS J10245338-5749040 are marked in the map.  
Right panel: Colour map of magnetic field strength from a 3D MHD simulation of the plasma flow produced by interacting winds from young massive stars in the cluster. The map shown is a cross section, for the rendering presentation see \citet{2022MNRAS.517.2818B}.}
    \label{fig:wd2-art}
\end{figure*}

\section{Observational signatures of CR sources at MeV energies}\label{sec:signatures}
Gamma-ray emission, if correctly interpreted, provides direct insight into the acceleration properties of the sources, especially concerning their maximum achievable energy.
This crucial step often starts with the identification of the emission mechanism between leptonic and hadronic, since the percentage of energy released in gamma rays is quite different between the two processes, and consequently the same detected photon energy could correspond to a very different parent particle energy in the two cases. The distinction is not always straightforward in the GeV-TeV energy range, and usually requires deep morphological analysis with auxiliary multi-wavelength data aimed at constraining the target for particle interactions. Nevertheless, the distinction could be largely improved using MeV observations, \giada{ as expain in the following subsections. }

\begin{figure}
    \centering
    \includegraphics[width=0.5\linewidth]{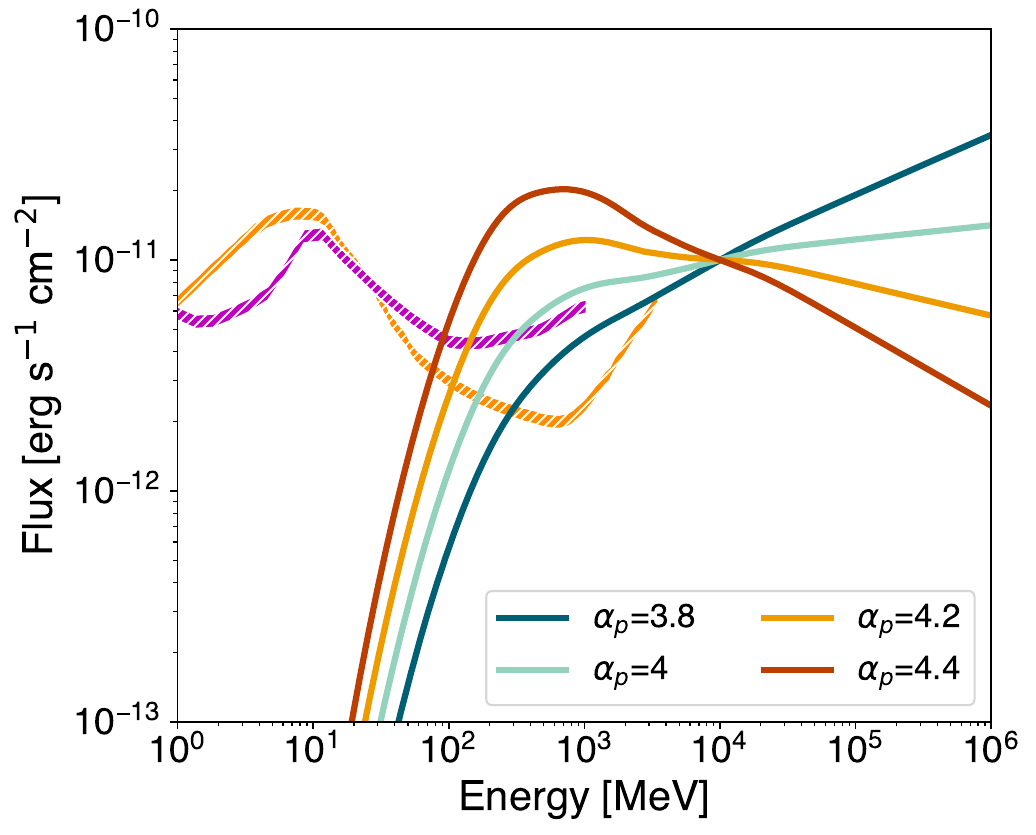}\includegraphics[width=0.5\linewidth]{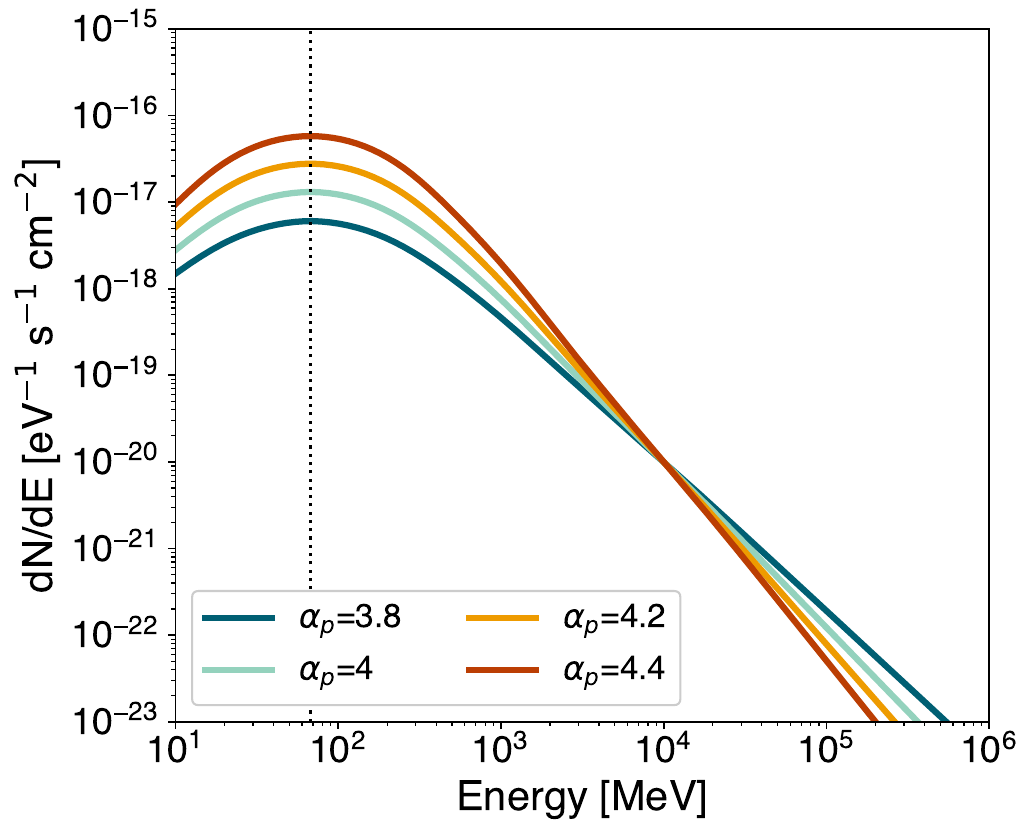}
    \caption{\red{Photon spectrum derived from proton-proton interactions assuming different spectra of the parent protons \ea{(with spectral index in momentum as indicated in the figure legend)}\giada{; the spectrum is normalized to the flux at 10 GeV measured from the SNR W44.}  \textit{Left panel}: photon flux in the $E_{\gamma}^2\; dN/d E_{\gamma}$ representation, showing how the appearance of a peak depends on the slope of the proton spectrum.  \giada{The sensitivities of 3-years observations with Amego-X and newASTROGAM are shown in magenta and orange respectively.} \textit{Right panel}: photon spectrum in the $ dN/d E_{\gamma}$ representation showing the typical pion bump at $67.5$ MeV \giada{(dotted line)}. }}
    \label{fig:pionbump}
\end{figure}

\subsection{Leptonic vs. hadronic scenario in sources}
\giada{Inelastic nuclear collisions that give birth to unstable mesons and subsequent gamma-ray radiation have a characteristic threshold, determined by the mass of the produced particles. For pion production the energy threshold is 280~MeV. Above this threshold,} gamma-ray photons have typical energies of $\sim$10\% of the energy of the parent protons and keep approximately the same spectral distribution. \giada{As a consequence,} gamma-ray emission \giada{of hadronic origin} is strongly suppressed below \giada{the photon energy that corresponds to this threshold.} \red{\giada{More precisely,} the resulting  gamma-ray spectrum in the $dN/d E_{\gamma}$ representation exhibits a typical feature called \textit{pion-bump} around half the mass of the neutral pion (${m_{\pi_0}}/2 = 67.5$ MeV). As shown in Fig.~\ref{fig:pionbump}, the position of the bump remains essentially unchanged for different proton spectral slopes.  
}
\red{
Such a characteristic bell-shaped feature can appear also in the $E_{\gamma}^2\; dN/d E_{\gamma}$ flux representation \giada{, more typically used for representation of SEDs; in this latter case the peak falls} at a characteristic photon energy of $\approx 1~$GeV \giada{and it is usually sampled with instruments like Fermi-LAT or AGILE.} \giada{Nevertheless, the shape of the peak and its exact location depend} on the slope of the parent proton spectrum, as shown in Fig.~\ref{fig:pionbump}, and it can eventually disappear for spectra harder than $\propto p^{-4.1}$. Therefore, one should be careful in interpreting a hard SED as leptonic emission, and, at the same time, one should be careful in interpreting a sub-GeV suppression as a smoking-gun evidence of hadronic emission.   }  

\red{
\giada{A well-studied example is provided by the SNR W44, for which a} a peak around 1 GeV is clearly visible \giada{in the $E^2$-representation (see Fig.~\ref{fig:w44}) \citep{agilew44,Fermi_pionbump,peron2020}. Likewise also in the SNRs IC433 \citep{Fermi_pionbump,IC433_Agile} and W51C \citep{Fermi_w51c} a similar feature is recorded}}.
\giada{This feature well matches with the expectation for hadronic emission.} However, as discussed by a few authors \citep{ambrogi2019,peron2020}, a similar spectral shape can in principle also be obtained in a leptonic scenario, e.g. through bremsstrahlung emission, if one allows for a low-energy break of the electron distribution. In the case of W44, a curved shape is possibly obtained from bremsstrahlung emission by suppressing the electron spectrum below 600 MeV \citep{peron2020}. This may appear as an extreme assumption, but such a spectrum is also invoked to better accommodate the synchrotron emission observed in the interstellar medium \cite{Orlando2018}. The gamma-ray emission, in the case of a break in the spectrum of the parent electrons, has a specific spectral slope, which starts to be distinguishable from the pion-decay shape only below 100 MeV, at the edge of Fermi-LAT observational window. Sensitive enough observations in that energy range would provide unequivocal evidence towards a leptonic or hadronic origin of the emission, allowing a huge step forward in our understanding of particle acceleration in different types of sources, even beyond SNRs.

\subsection{Electron-to-proton ratio}
Another open question in the theory of particle acceleration concerns the fraction of leptons that are accelerated along with protons and other nuclei. From theory and simulations a very small fraction ($K_{ep} < 0.01 $ \cite{Merten2017, Cristofari2021}) of electrons is expected to be accelerated compared to protons in the context of stochastic acceleration (see discussion in Sec. \ref{sec:acc_trans}). A few studies constrained this quantity for SNRs, finding consistent values (see e.g. \cite{morlino_caprioli}). However, also in this case, this type of measurements requires a broad-band modeling of data collected in different wavebands. This joint information is not always available and is often subject to degeneracy between different parameters, as one is required to model the pion emission, that depends on the gas density and on the number of nuclei, along with the synchrotron emission, which in turn depends on the magnetic field and on the number of electrons. In addition, being synchrotron radiation and pion emission probed by different experiments, one can hardly find the two components being completely co-spatial. The availability of MeV observations, instead, would allow us to directly probe the fraction of electrons-to-protons accelerated in a source. The MeV range, in fact, has the great advantage of hosting the transition between pion and bremsstrahlung emission, with the former dominating above a few tens of MeV and the latter below. Both processes are proportional to the target column density, therefore their ratio is directly connected to the ratio of electrons to protons. The right panel of Fig.\ref{fig:w44} shows the prediction for the bremsstrahlung component in W44, assuming different values of $K_{ep}$ and the same injection slope for electrons and protons. The two components are well separated in the MeV regime, where the future MeV detectors will operate. This will provide strong constraints on the $K_{ep}$ ratio.   
 
\begin{figure}
    \centering
    \includegraphics[width=0.5\linewidth]{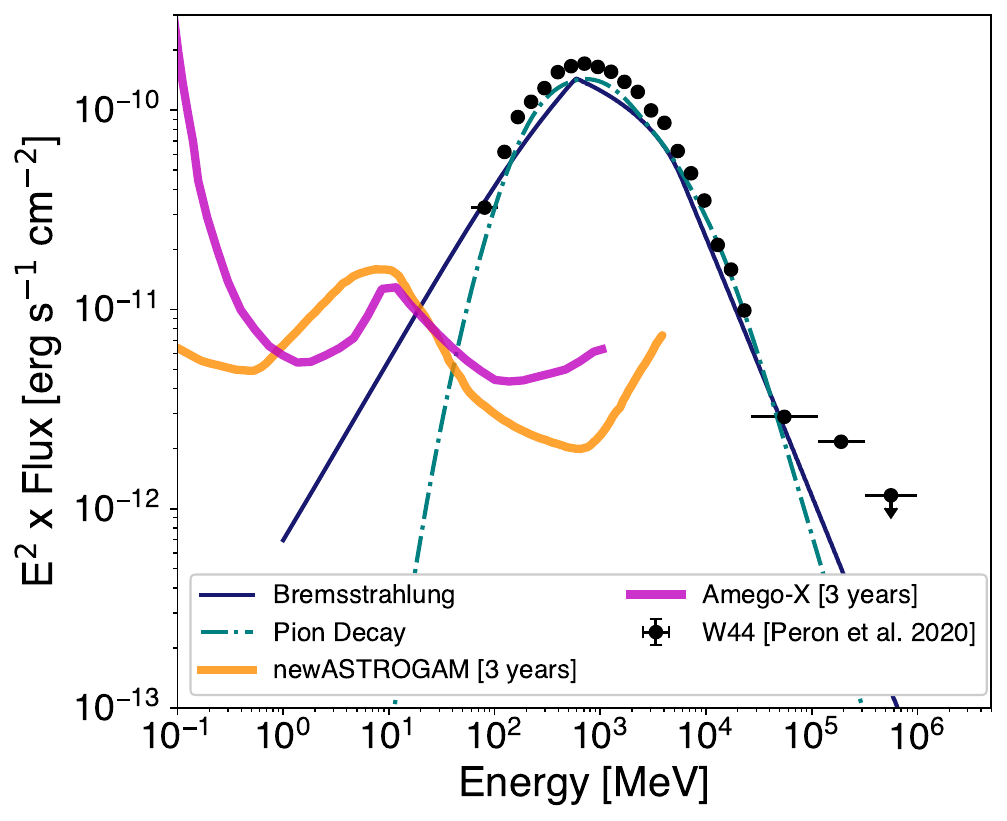}\includegraphics[width=0.5\linewidth]{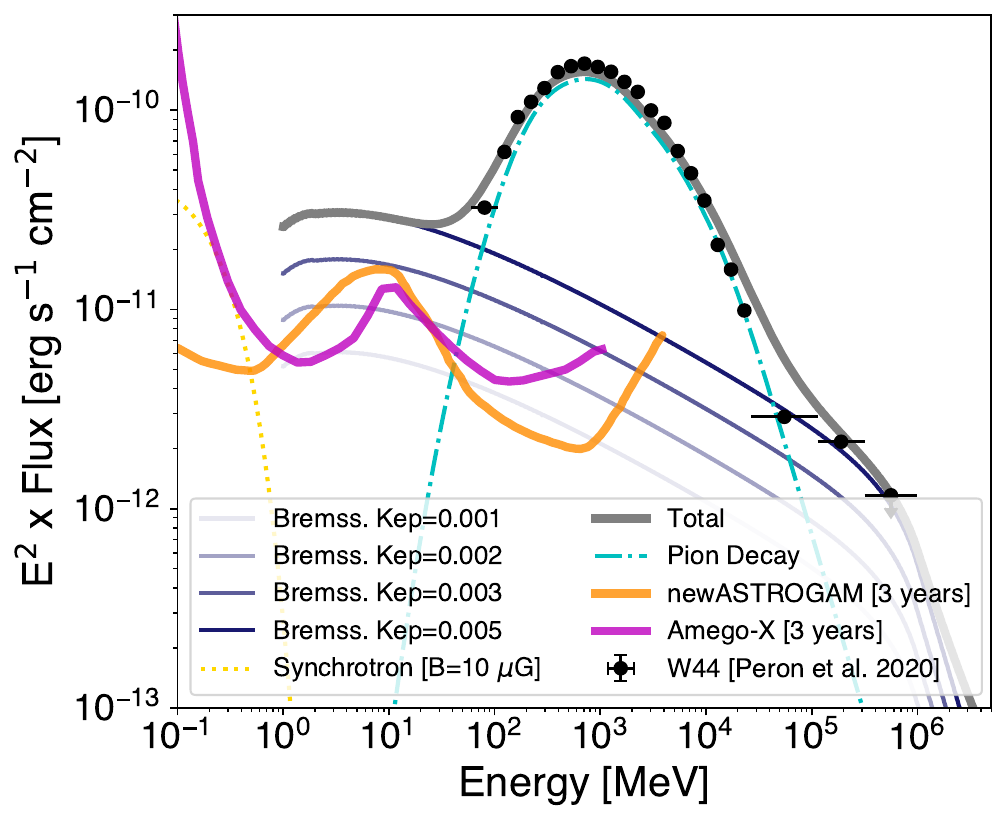}
    \caption{Modelling of the SED derived from Fermi-LAT observations of the middle-aged supernova remnant W44 \cite{peron2020}. On the left, two possible scenarios are compared as the origin of the emission: hadronic (pion decay) or leptonic (bremsstrahlung). On the right, the emission is modelled with a dominant hadronic component, and a sub-dominant leptonic component, which derives from bremsstrahlung of electrons. The normalization of the electron population varies with different $K_{ep}$, while the injection slope is the same as for protons. The sensitivity curves of newASTROGAM and Amego-X are the same as in Fig. \ref{fig:snr_cat} }
    \label{fig:w44}
\end{figure}

\subsection{Synchrotron and bremsstrahlung emission from secondary electrons and acceleration efficiency}
Hadronic interactions produce, along with gamma-ray photons, also electrons and positrons, that, in turn, produce leptonic radiation. The electrons generated as secondary products of proton-proton interactions take on approximately 5\% of the energy of the primaries and produce emission via bremsstrahlung,\red{inverse Compton,} and synchrotron processes. While the former \red{two contributions are} usually subdominant in hadronic-dominated cases \red{in the MeV regime, because the spectrum of secondary particles will fall in the same range as the pion emission}, the synchrotron component generated by secondary electrons could emerge as a signature that pinpoints the presence of higher energy primaries. If the energy of the primary nuclei and the intensity of the magnetic field are high enough (see extended discussion in \cite{celli2020}), this emission could be detected by future MeV detectors. This is particularly interesting for the investigation of hadronic PeVatrons, where large ($>> 1$mG) magnetic fields are needed to ensure acceleration to high enough energies. An example of the possible synchrotron emission deriving from secondary electrons is shown in Fig. \ref{fig:cygnus}, where emission from the Cygnus region is considered as an example. We assumed the high-energy emission from GeV to PeV to be predominantly hadronic and computed the synchrotron emission with different assumptions on the magnetic field. In principle, one should evaluate what the dominant timescales for losses are in the considered system. Here we assume that synchrotron losses dominate the cooling, due to the large assumed B, but a careful evaluation should be carried out. The additional information that could be derived by detecting secondary electrons would be extremely useful to characterize extreme accelerators, especially as it has the potential of constraining the magnetic field in the acceleration region. As demonstrated in Fig. \ref{fig:cygnus} the emission from secondary electrons arising from strong hadronic sources is within the reach of sensitivity of next-generation of MeV telescopes.
Moreover, the analysis of the emission of secondary electrons can also unveil cases of reacceleration. Having a clear understanding of the process is fundamental to assess the effective efficiency of a source, which would significantly be reduced in case of reacceleration of CRs \red{from} the Galactic pool. This issue was raised for example in the modeling of W44 \cite{Uchiyama2010,Lee2015,Cardillo2016} where a clear discrimination between scenarios invoking acceleration of fresh particles alone and a mix of acceleration and reacceleration was shown to rely on our ability to detect potential bremsstrahlung emission from secondaries at tens to hundreds of MeV energies (see e.g. Fig.3 and 4 of \cite{Cardillo2016}). For the prototypical case of W44, the expected flux in this energy range is $\gtrsim 3 \times 10^{-12} {\rm erg}\ {\rm s}^{-1} {\rm cm}^{-2}$ above 10 MeV, and hence detectable in principle by next generation MeV telescopes within the nominal 3-years exposure, putting these experiments in an ideal position to finally answer the question of acceleration efficiency in SNRs that are well established hadronic CR accelerators.

\begin{figure}
    \centering
    \includegraphics[width=1\linewidth]{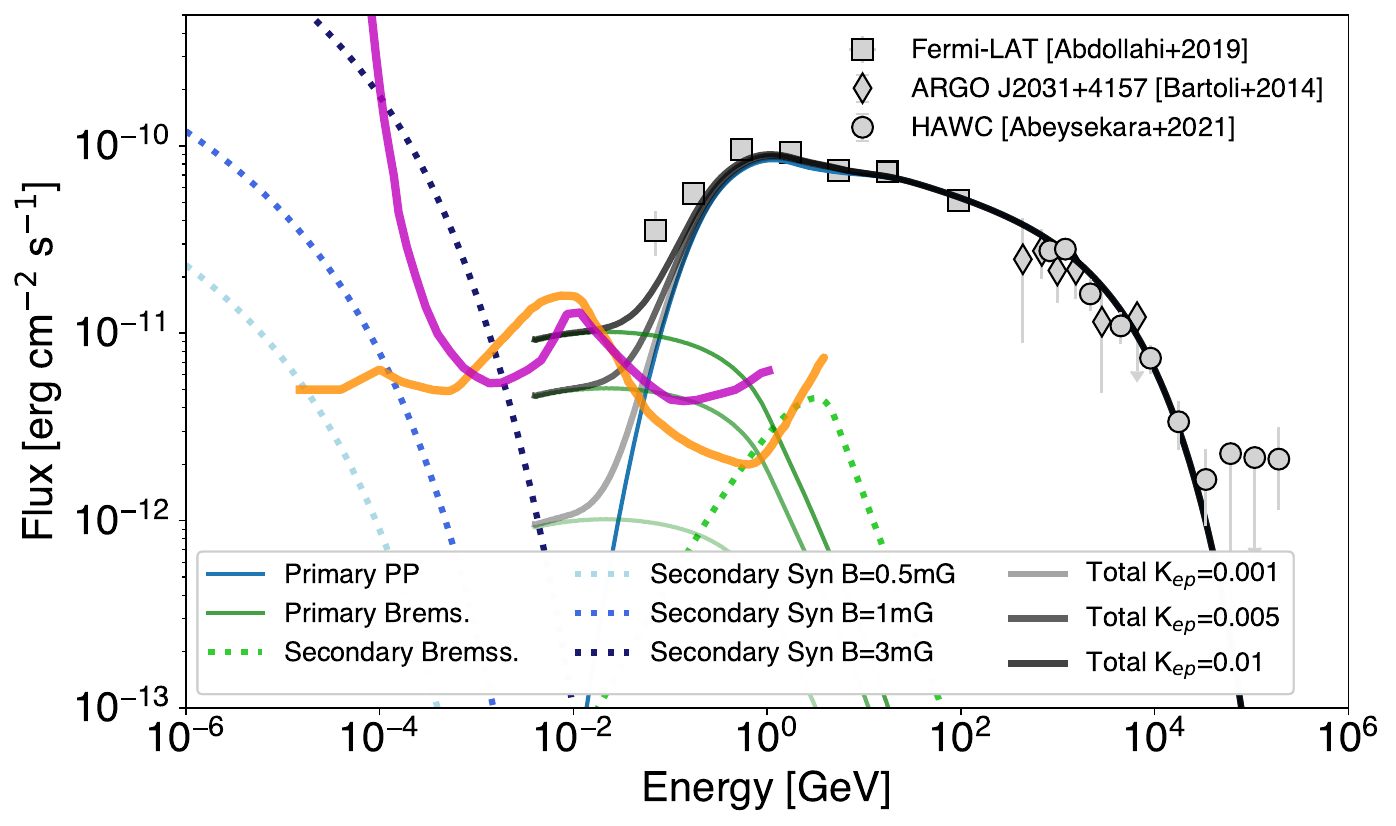}
    \caption{Broadband spectral modeling of the Cygnus cocoon. Data are collected by Fermi-LAT, ARGO, and HAWC as indicated in the figure label and refer to the compilation presented in \citep{Menchiari_2024}. The high-energy emission is modeled {as primarily due to} pion decay, assuming a target gas density of 10 cm$^{-3}$.  The relative contribution of primary bremsstrahlung (assuming different values of $K_{ep}$) is indicated in different shades of green. The secondary bremsstrahlung and synchrotron components are indicated as dotted lines. The latter has been computed for different values of the magnetic field, as indicated in the figure legend. {The orange and magenta curve are the sensitivity curves for 3-years observation with newASTROGAM and Amego-X, respectively \cite{Amego2022,BergeICRC}.}} 
    \label{fig:cygnus}
\end{figure}

\section{Probing MeV cosmic rays via their interactions with the interstellar medium}\label{sec:ism_signatures}
CRs, especially the lowest energy ones (MeV-GeV energies), have a deep impact on the interstellar medium, as they regulate the ionization state of the gas and therefore influence important processes such as star and planet formation \citep{2020SSRv..216...29P}. At the same time, they leave signatures that one can detect both in the form of {non-thermal continuum} emission and in the form of gamma-ray nuclear lines. In the following, we describe their impact on the ionization state of the interstellar medium and the observational features that they produce, while \red{we refer to}  \citep{SiegertISSI} for a comprehensive review on nucleosynthesis and the relative relevance of MeV observations. 
 
\subsection{Ionization of molecular clouds}
\label{subsec:MCIonization}
Differently from ionizing radiation \citep{McKee_XrayAbsorption_1989}, low-energy CRs ($E\lesssim 1$ GeV) can penetrate deep within the core of gas clumps, producing ionization of molecular clouds (MCs), possibly heating the dense cold gas \citep{Galli_CRHating_2015}, and indirectly inducing chemical reactions in the interstellar medium (ISM), generating complex molecular compounds \citep{Dalgarno_CRsISMChem_2006}. An exhaustive investigation of the ionization processes in MCs and diffuse atomic clouds can be found in \cite{Padovani_MCIon_2009}. A plethora of interactions can lead to the ionization of $H_2$\footnote{MCs also harbor a fraction of helium, which can be likewise ionized (see \cite{Padovani_MCIon_2009})}, namely, proton-induced ionization\footnote{Also nuclei may induce ionization but, for the sake of simplicity, we will only consider ionization generated by protons.}:

\begin{subequations}
\begin{gather}
p_{CR}+H_2 \rightarrow p_{CR} + H_2^+ + e \label{eq:pDirIonReac} \\
p_{CR}+H_2 \rightarrow  H + H_2^+ \label{eq:pElCaptIonReac} \\
p_{CR}+H_2 \rightarrow p_{CR} + H + H^+ + e \label{eq:pDissIonReac} \\
p_{CR}+H_2 \rightarrow p_{CR} + 2H^+ + 2 e \label{eq:pDoubleIonReac}
\end{gather}
\end{subequations}
or ionization induced by CR electrons:
\begin{subequations}
\begin{gather}
e_{CR}+H_2 \rightarrow e_{CR} + H_2^+ + e \label{eq:eDirIonReac} \\
e_{CR}+H_2 \rightarrow  e_{CR} + H + H^+ + e \label{eq:eDissIonReac} \\
e_{CR}+H_2 \rightarrow e_{CR} + 2H^+ + 2 e \ . \label{eq:eDoubleIonReac}
\end{gather}
\end{subequations}

The molecular hydrogen ion ($H_2^+$) production rate from the sole contribution of protons and electrons, through \textit{direct ionization} (Eq. \ref{eq:pDirIonReac} and Eq. \ref{eq:eDirIonReac}) and \textit{electron capture} (Eq. \ref{eq:pElCaptIonReac}) processes can be calculated as:
\begin{equation}
\label{eq:IonRateIons}
\zeta_{H2}= \sum_k \int_{I(H_2)}^{E_{max}} c f_k(E_k) [1 + \phi_k(E_k)] \sigma_k^{\rm ion}(E_k) dE_k  + 
\int_0^{E_{max}} c f_p(E_p) \sigma_p^{\rm e.c.}(E_p) dE_p
\end{equation}
\red{where the index $k$ accounts} for the considered CR species (electrons or protons), $I(H_2)=15.603$ eV is the ionization potential of H$_2$, and $\sigma^{\rm ion}$ and $\sigma^{\rm e.c}$ are the direct ionization and electron capture cross sections respectively (see \cite{Padovani_MCIon_2009} and references therein). The quantity $\phi_k(E_k)$ is a correction factor accounting for the ionization induced by a population of secondary electrons created by direct ionization, and can be calculated as: 
\begin{equation}
\phi_k(E_k)=\frac{1}{\sigma_k^{\rm ion}(E_k)} \int_{I(H_2)}^{E_{max}} \mathcal{P}(E_k, E'_e) \sigma^{\rm ion}_e(E'_e)dE'_e
\end{equation}
with the term $\mathcal{P}(E_k, E'_e)$ describing the probability that a secondary electron with energy E'$_e$ is created during a primary ionization by a particle with energy E$_k$. 

\red{Figure~\ref{fig:SigmaIon} (left panel) shows the ionisation cross section, $\sigma^{\rm ion}$, for protons and electrons ($k = p, e$). The cross section for protons (electrons) peaks at about 100~MeV (100~keV) and then rapidly decreases, becoming negligible compared to other interaction channels at GeV energies \citep{padovani2020}. 
This behavior implies that, if the proton and electron spectra measured by \textit{Voyager} are representative of the Galactic LECR population, the dominant contribution to ionisation by protons arises from particles with kinetic energies around $\sim 100$~MeV.
This is evident from the differential ionisation rate, $E_k,d\zeta/dE_k$ (Eq.~\ref{eq:IonRateIons}), which peaks near this energy, as shown in Fig.~\ref{fig:SigmaIon} (right panel). In contrast, the same figure shows that no such peak appears in the electron case within the energy range probed by \textit{Voyager}. Consequently, \textit{Voyager} data do not currently constrain the ionisation rate contributed by CR electrons. Notice that different LECR spectra would lead to different conclusions. 
}

\red{In principle, electrons with energies below the \textit{Voyager} range, or, more generally, additional hidden low-energy components (the so-called CR \textit{“carrots”})—could significantly enhance the ionisation rate in molecular clouds. Nevertheless, such scenarios, sometimes proposed in the literature, are expected to face severe energetic constraints (see \cite{recchia2019-carrot} for a dedicated discussion).
}

Creation of $H_2^+$ ions in a dense environment can trigger an intricate chain of chemical reactions with the formation of complex molecules, such as, for example, DCO$^+$ and HCO$^+$ ($D$ stands for the deuterium isotope). The detection of these species through molecular emission lines in the radio can be used to assess the value of $\zeta_{H2}$ \citep{Caselli_IonFracMC_1998, Vaupre_W28IonRate_2014}. In a steady-state regime, the abundances of DCO$^+$ and HCO$^+$ are set by the following main reactions \citep{Guelin_DCO+_1977, Caselli_IonFracMC_1998}:
\begin{subequations}
\label{eq:ChemNet}
\begin{gather}
k_{CR}+H_2 \xrightarrow[]{\zeta_{H2}} k_{CR} + H_2^+ + e \\
H_2^++H_2 \xrightarrow[]{\kappa_{H_2^+}}  H_3^+ + H \\
H_3^++CO \xrightarrow[]{\kappa_H} HCO^+ + H_2 \\
HCO^+ + e \xrightarrow[]{\beta'} CO + H \\
H_3^+ + e \xrightarrow[]{\beta} 3H \text{ (or $H_2+H$)}\\
H+H \xrightarrow[]{\kappa'} H_2 \\
H_3^++HD \xrightleftharpoons[\kappa_f^{-1}]{\kappa_f} H_2D^+ + H_2  \\
H_2D^+ + CO \xrightarrow[]{\kappa_D} DCO^+ + H_2 \\
DCO^+ + e \xrightarrow[]{\beta'} CO+D  \\
H_2D^++e \xrightarrow[]{\kappa_e} 2H+D \text{ (or $H_2+D$ or $HD+H$)} \\
H+D \xrightarrow[]{\kappa''} HD \\
H_2D^+ + CO \xrightarrow[]{\kappa'_D} HCO^+ + H_2\\
H_3^+ + D \xrightleftharpoons[\kappa_f'^{-1}]{\kappa_f'} H_2D^+ + H  \\
CO^+ + HD \xrightarrow[]{\kappa_{CO^+}} DCO^+ + H
\end{gather}
\end{subequations}
where the parameters on top of the arrows denote the creation (or destruction) rates of each chemical compound (see \cite{Vaupre_W28IonRate_2014}). The ionization rate $\zeta_{H2}$ can then be analytically expressed in terms of the abundance ratios $R_D=[DCO^+]/[HCO^+]$ and $R_H=[HCO^+]/[CO]$ \citep{Wootten_ElecAbundance_1979, Guelin_IonStateMC_1982}:
\begin{equation}
R_D=\frac{[DCO^+]}{[HCO^+]}\simeq \frac{1}{3} \frac{x(H_2D^+)}{x(H_3^+)} \simeq
\frac{1}{3} \frac{\kappa_f x(HD)}{\kappa_e+x(e)+\delta+\kappa^{-1}_f/2}
\end{equation}
\begin{equation}
\label{eq:RatHCO+CO}
R_H=\frac{[HCO^+]}{[CO]} = \frac{\kappa_H x(H_3^+)}{\beta' x(e)} \simeq
 \frac{\kappa_H}{x(e)[2 \beta x(e)+ \delta]\beta'} \frac{\zeta_{H2}}{n_H}
\end{equation}
where $\kappa_f$, $\kappa_e$, $\kappa_H$, $\beta$ and $\beta'$ are the reaction rates\footnote{Note that the reaction coefficients actually mediate collisional processes, and, as a consequence, some of them depend on the kinetic temperature of the gas \citep{Caselli_IonFracMC_1998}.} for the different processes within the chemical networks of Eq.\ref{eq:ChemNet}, $\delta \approx \delta_{H_3^+} \approx \delta_{H_2D^+}$ is the total destruction rate of H$_3^+$ or H$_2$D$^+$ due to reactions with neutral species such as CO and O, and $x(\mathcal{X})=n_\mathcal{X}/n_H$ denotes the fractional abundance of a given specie $\mathcal{X}$ with number density $n_\mathcal{X}$. 

Inversion of Eq. \ref{eq:RatHCO+CO} gives the expression for $\zeta_{H2}$ \citep{Caselli_IonFracMC_1998}:
\begin{equation}
\label{eq:IonRate}
\zeta_{H2}= [2 \beta x(e)+ \delta]\frac{R_H x(e) \beta' n_H}{\kappa_H} \simeq \left[ 7.5 \times 10^{-4} x(e) + \frac{4.6 \times 10^{-10}}{f_D} \right] x(e) n_H R_H
\end{equation}
where $f_D$ is the depletion factor of C and O, defined such that $1/f_D$ is the fraction of C and O in the gas phase (note that $f_D$ is the same for C and O \citep{LMH_1984}), and
\begin{equation}
x(e)= \left[ \frac{\kappa_f x(HD)}{3 R_D} - \delta \right] \simeq \frac{2.7 \times 10^{-8}}{R_D} - \frac{1.2 \times 10^{-6}}{f_D}\ .
\end{equation}

\begin{figure*}[t]
\centering
\includegraphics[width=0.5\linewidth]{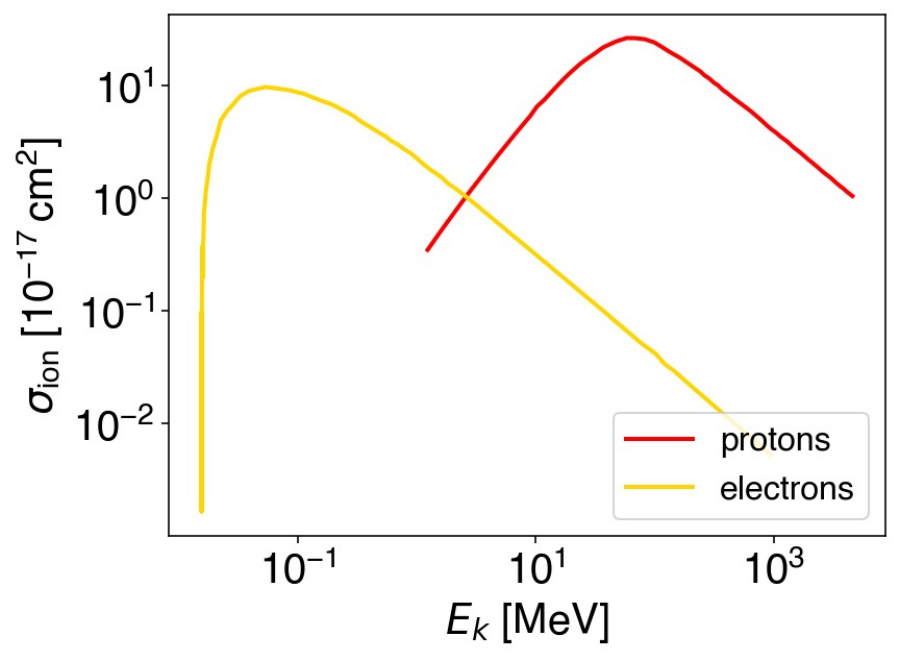}\includegraphics[width=0.5\linewidth]{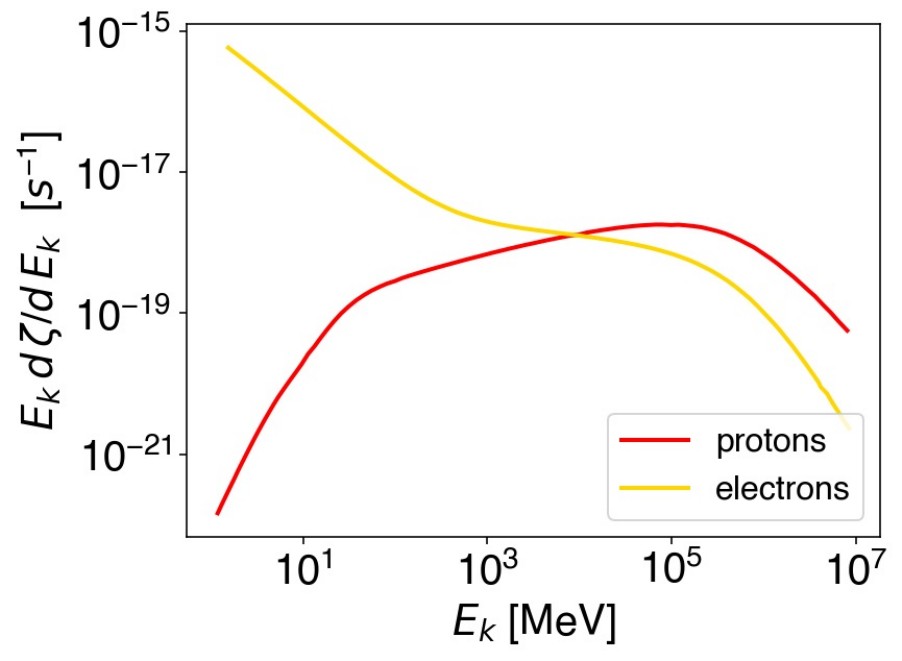}
\caption{\red{Left panel: Direct ionisation cross section of protons and electrons ($k=p, e$) as a function of the kinetic energy \protect\cite{Padovani_MCIon_2009}; Right panel:    Differential ionization rate of protons and electrons corresponding to Voyager spectra \protect\cite{Phan-2018-ionization-MC}.}}
\label{fig:SigmaIon}
\end{figure*}

Assuming we know the depletion factor, $\zeta_{H2}$ is easily obtained as $R_D$ and $R_H$ can be readily estimated from observations through the ratios of molecular rotational lines. The approach reported by \cite{Caselli_IonFracMC_1998} represents pioneering work in the development of analytical methods for the estimation of the ionization rate. However, this method comes with some caveats, and it is not generally a solid approach. In fact, in regions with a high degree of CO depletion, together with the possible existence of sharp density and temperature gradients in the MCs, makes Eq.~\ref{eq:IonRate} no longer valid \citep{Caselli_DeutProbIon_2002}.
The state of the art analytical approaches for the estimation of the ionisation rate are presented in \cite{Bovino_CRionDenseCore_2020}, where the ionisation rate is derived from observational quantities and the method is validated using 3D numerical simulations.

As a final remark, it is essential to emphasize that the value of $\zeta_{H2}$ does not return any information on the spectral shape of the CR distribution since the spectral information is lost in the integral in Eq. \ref{eq:IonRateIons}. {In this context, MeV observations may play a crucial role, as} one could {derive the CR spectrum from low energy gamma-ray emission, compute the expected ionization rate and use the observed value as a cross-check} to have a comprehensively self-consistent estimate of the CR level at different locations. \red{We also note that secondary electrons and positrons may contribute to the ionization rate, thereby influencing the sensitivity of this method to hadronic CRs, although their actual relevance must be assessed on a case-by-case basis.}
{Only a very limited number of studies have so far attempted to connect the measure of the ionization rates with gamma-ray emission \citep{{phan2020}}. Extending these studies would be very insightful for the modeling of LECRs. }

{Several measurements of the CR ionization rate have been obtained exploiting the information derived from the typical spectral lines of $H^{+}_{3}$, OH$^{-}$ and other ions observed in the infrared regime (see Figure \ref{fig:ionization_phan}). A broad discussion on the constraints on the CR spectrum that can be obtained from these data is reported in \cite{gabici2022}. The average ionization rate in the local diffuse medium appears to be largely enhanced, by more than one order of magnitude, compared to the ionization rate that could be induced by the locally measured CR spectrum, namely $1.51 - 1.64 \times 10^{-17}$ s$^{-1}$. However, a recent re-evaluation \citep{Obolentseva2024} of the gas column density reduced the discrepancy between the two quantities {in the local  (within a few kpc from us) environment.} 
A large value of the ionization rate is {measured in the Galactic center region}, $\zeta \approx (2-7) \times 10^{-15}$ s$^{-1}$, possibly caused by the several CR accelerators, including some of the most massive star clusters in the Galaxy, like the Arches and the Quintuplet clusters, located in the central molecular zone. However, a recent study \citep{Ravikularaman-2025-ionization-GC} concluded that it is hard to reconcile the gamma-ray emission, naturally connected to CRs, with such a large ionization rate, suggesting that more factors beyond CRs are influencing the ionization state in the region. 

\begin{figure}
    \centering
    \includegraphics[width=0.8\linewidth]{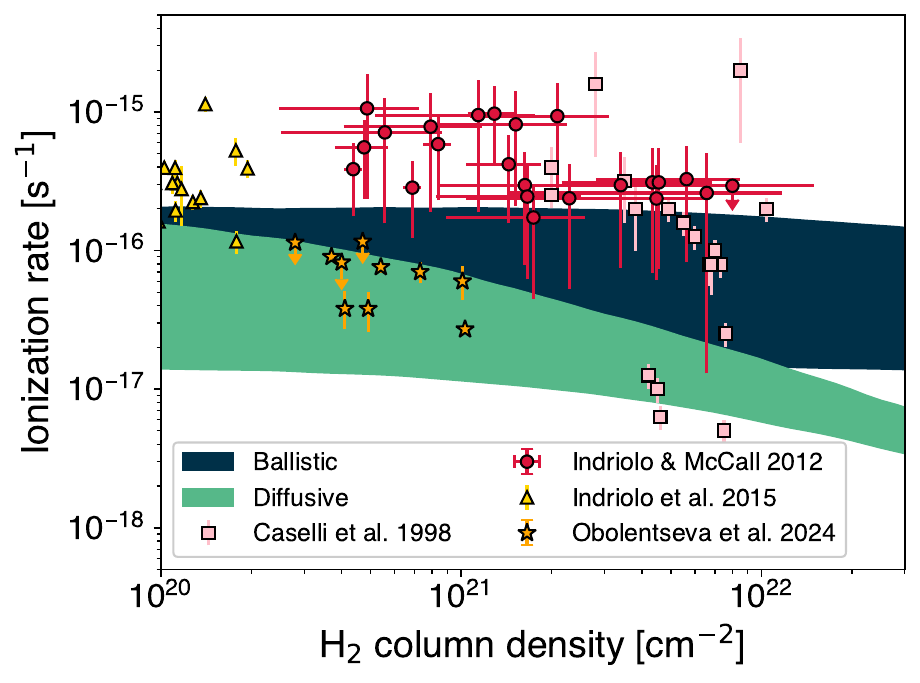}
    \caption{A compilation of measurements of the ionization rate as derived by different tracers along lines of sight of different gas column density. For comparison, the local expectations based on measured CRs and assuming a certain degree of stochasticisty of nearby sources, are reported in green and blue, and are relative to a diffusive (green) or a ballistic (black) propagation regime of particles. The figure has been adapted from \citep{Phan2023}. }
    \label{fig:ionization_phan}
\end{figure}

\subsection{Gamma-ray emission from molecular clouds}\label{subsec:MC}
During their propagation through the Galaxy, CRs leave imprints in the ISM that they cross, by interacting with the gas and producing secondary emission such as gamma rays. The latter carry information on the spectrum of the parent CRs that have generated them. If the target density is high enough, the gamma-ray emission can be detected and used as a tracer of the CR population far from their accelerators. This possibility has been extensively used in the gamma-ray community, especially using Fermi-LAT data aimed at analyzing both the large scale diffuse emission \cite{Fermi_diffuse2012,Fermi_diffuse2015,Fermi_diffuse2016,Yang_diffuse2016,Pothast_2018}, and compact molecular clouds \cite{Yang_nearby2015,Yang_sgr,aharonian2020,Baghmanyan2020,peron2021}. 

The analysis of the diffuse emission detected by \textit{Fermi}-LAT unveiled some inhomogeneity in the gamma-ray emissivity of the gas in the inner region ($\sim$ 2-5~kpc) of the Milky Way \citep{Fermi_diffuse2016,Yang_diffuse2016, Pothast_2018}. This is often interpreted as the result of the clustering of sources in the inner part of the Galaxy. {However a number of different explanations have been proposed, invoking either \textit{intrinsic} or \textit{extrinsic} effects. Examples of the former type are an enhanced density of freshly injected particles in regions that are crowded with sources \citep{Recchia2016diffuse}, or a non-uniform propagation of particles in the inner Galaxy, influenced by the enhanced turbulence in the vicinity of accelerators \citep{Tibaldo_gaggero_review}. A different scenario is that associating the enhanced emissivity to the superposition of faint, unresolved gamma-ray sources along the line of sight \cite{Vecchiotti2022Fermi,Pagliaroli2023}.} 

{This ambiguity should be reduced when one focuses on the gamma-ray emission from compact regions of large target density, such as molecular clouds.} Analysis of molecular clouds in the \textit{Fermi}-LAT waveband \citep{aharonian2020} emphasized a similar trend also for clouds, with a tendency to {reveal} enhanced emissivities in the inner part of the Galaxy, but with important exceptions: indeed the emissivity derived for a few clouds in the inner Galaxy {turns out to be much lower than that of the diffuse medium in the same region }, suggesting that the enhancement  is likely produced on a smaller scale and associated with an increased abundance of CR accelerators \cite{peron2021}. 

In the medium within a few kpc from us, the CR spectrum derived from the gamma-ray emissivity of {the diffuse medium} and clouds at GeV energies seems to be compatible, within a factor $\lesssim$2, with expectations based on the local CR spectrum as traced by Voyager, AMS-02, and other instruments \cite{Orlando2018}. This is {not surprising}, because the Fermi-LAT band traces CRs of energies beyond a few GeV, where no large fluctuations are foreseen. The situation is different at lower energies, as discussed in section \ref{subsec:lecr_transport} and shown in Fig. \ref{fig:stochasticity-proton}. The regime where the fluctuations are expected to be largest is not directly accessible by gamma-ray observations due to the threshold for pion production at 280 MeV. Below that energy information is only available from measurements of ionization rates and nuclear lines. On the other hand, also in the GeV band a few deviations from the Voyager expectations emerged \cite{Yang2023EffectiveClumps} in some clumps of high density in the Taurus molecular cloud. This result may indicate that propagation of CRs may be strongly influenced by the phase of the ISM as suggested in \cite{Chernyshov2024Self-consistentCase}.

Gamma-ray observations can be used as complementary information to investigate and constrain the excitation mechanism. While Fermi-LAT observations are challenging in these energy range, future MeV detectors as newASTROGAM and AMEGO-X could provide insightful observations towards molecular clouds and investigate the spectrum of CRs in their location, testing for possible deviations from the local CR spectrum. As shown in Fig. \ref{fig:clouds_mev}, despite the low surface brightness that characterizes molecular clouds, future MeV instruments {will be sensitive enough to detect} a large fraction of them. The visibility is determined by comparing the expected flux of a cloud, with the expected sensitivity of future MeV instruments. The cloud flux depends only on the ratio between its mass and the square of its distance, once the illuminating CR spectrum is fixed. In Fig. \ref{fig:clouds_mev} we consider the local CR spectrum measured at Earth and a range of values for the ratio $M/d^{2}$ normalized to the values of $10^5 $M$_\odot$ and 1~kpc. As the clouds are not point-like sources, the sensitivity {must be worsened} to take into account the additional background that affects extended sources: {this is done roughly by multiplying the point source sensitivity by a factor $\propto \sqrt{\theta^2_{src}+\sigma^2_{PSF}}$}. 
Even accounting for {an extension as large as} 1 degree (dotted curves in the plot), clouds with a broad range of parameters are in the reach of future MeV missions. 

\subsection{MeV de-excitation lines}
It is worth mentioning that MeV regime gamma-ray lines from LECRs interactions with ambient matter \citep{1979ApJS...40..487R,1996ApJ...472..205T,2003EAS.....7...79T} 
can be very informative tools to study both CR acceleration processes and the chemical composition of the ISM. These lines fall in the MeV range and are generated by CRs of energies of $\sim$ 1-100 MeV/nucleon. Estimations of expected gamma-ray line fluxes from CR sources such as massive star associations and clusters \citep{2014A&ARv..22...77B} suggest values of $\sim$ 10$^{-14}$ \fluxcgs; this may be enough to detect the lines with future instruments. Clusters embedding young supernova remnants could have a LECR population with a composition enriched in metals which were accelerated from material ejected by core-collapse supernovae. 
Meanwhile, measurements of MeV lines from the diffuse medium are a promising tool for investigating the spectral properties of LECRs. We refer to \citep{SiegertISSI} for more details regarding the production and the detection prospects of gamma-ray lines in the MeV band. 

\begin{figure}
    \centering
    \includegraphics[width=0.5\linewidth]{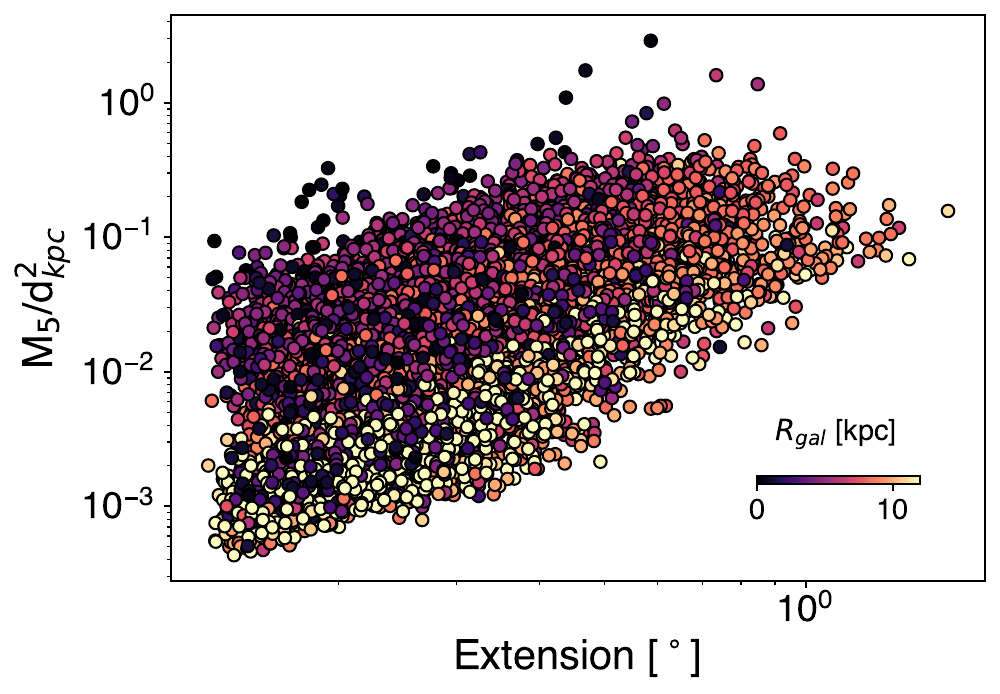}\includegraphics[width=0.5\linewidth]{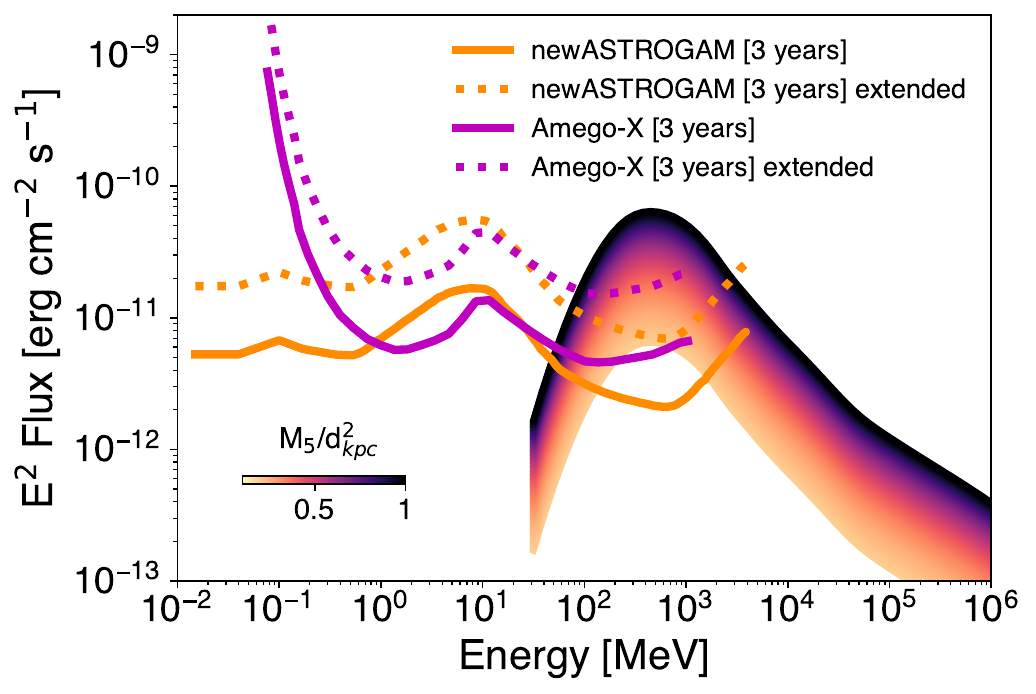}
    \caption{On the left, the distribution of the factor $M/d^2$ (normalized to $M_5=M/10^{5}~\rm{M_{\odot}}$ and to $d_{kpc}=d/1~kpc$) as a function of the extension for the sample of molecular clouds detected in the Milky Way \citep{Miville-Deschenes2016}. The color scale represents their distance from the Galactic center. On the right, the expected emissivity of a cloud that emits as illuminated by the local CR spectrum is shown for various values of $M_5/d_{kpc}^2$ and is compared with the sensitivity of future MeV detectors for point-like and extended sources (see text) \citep{Amego2022,BergeICRC}. }
    \label{fig:clouds_mev}
\end{figure}

\section{Summary}
In this article, we have reviewed \red{some of the} main properties of Galactic CRs, with a specific focus on LECRs. This component is an important regulator of the physics of the interstellar gas, mainly because of its ionizing power. Still, it is not very well constrained: direct measurements are challenged by solar modulation, while indirect measurements are challenged by the lack of instrumentation. The advent of future MeV detectors, like newASTROGAM and Amego-X, promises to provide an adequate sensitivity to probe CR signatures at low energies and finally constrain their flux intensity. Measurements towards the diffuse emission and towards clouds, both of continuous gamma-ray emission, and of gamma-ray lines, will provide direct insight into the abundance of LECRs, which can be compared with the available measurements of the ionization rate derived from infrared lines. Furthermore, data collected at MeV energies towards Galactic CR accelerators, like SNRs and SCs, will help unveiling the acceleration mechanisms and their energetics, as they will help constraining i) the nature of non-thermal high-energy emission (whether hadronic or leptonic), ii) the ratio between accelerated electrons and protons, iii) the amount of high energy protons, from the observations of low-energy secondaries, and iv) potentially discover new sources or source classes. 

In summary, upcoming MeV observatories hold the promise to greatly improve our knowledge of CR physics, and of LECRs in particular, with enormous implications on our understanding of the Galactic ecosystem.

\section*{Declarations}
\textbf{Competing Interests and Fundings:} No specific funding was received to assist with the preparation of this manuscript. The authors have no competing interests to declare that are relevant to the content of this article.

\bibliography{sn-article}

\end{document}